\documentclass[twocolumn]{aastex63} 

\usepackage[T1]{fontenc} 
\usepackage{courier}
\usepackage{amsmath}
\usepackage{enumerate}
\usepackage{tabularx}
\usepackage{multirow}
\usepackage{appendix}
\usepackage{xcolor} 
\hypersetup{linkcolor=magenta}   
\usepackage{comment}

\usepackage{dutchcal}
\usepackage{float}
\usepackage[figuresright]{rotating}

\begin{document}

\shorttitle{Probing Axion-Photon Mixing with High-Resolution X-ray Spectroscopy}

\title{Probing Axion-Photon Mixing with High-Resolution X-ray Spectroscopy}

\author[0000-0002-5793-554X]{Yu Zhou}
\affil{National Astronomical Observatories, Chinese Academy of Sciences (NAOC), 20A Datun Road, Chaoyang District, Beijing 100101, China}
\affil{International Center for Quantum-field Measurement Systems for Studies of the Universe and Particles (QUP, WPI),
High Energy Accelerator Research Organization (KEK), Oho 1-1, Tsukuba, Ibaraki 305-0801, Japan}

\author{Jiejia Liu}
\affil{Department of Astronomy, Tsinghua University, Beijing 100084, China}

\author{Volodymyr Takhistov}
\affil{International Center for Quantum-field Measurement Systems for Studies of the Universe and Particles (QUP, WPI),
High Energy Accelerator Research Organization (KEK), Oho 1-1, Tsukuba, Ibaraki 305-0801, Japan}
\affil{Theory Center, Institute of Particle and Nuclear Studies (IPNS),  High Energy Accelerator Research Organization (KEK), Tsukuba 305-0801, Japan}
\affil{Graduate University for Advanced Studies (SOKENDAI), 
1-1 Oho, Tsukuba, Ibaraki 305-0801, Japan}
\affil{Kavli Institute for the Physics and Mathematics of the Universe (WPI),   The University of Tokyo Institutes for Advanced Study, The University of Tokyo,  Kashiwa, Chiba 277-8583, Japan}

\author{Kazuhisa Mitsuda}
\affil{International Center for Quantum-field Measurement Systems for Studies of the Universe and Particles (QUP, WPI),
High Energy Accelerator Research Organization (KEK), Oho 1-1, Tsukuba, Ibaraki 305-0801, Japan}

\email{zhouyu@nao.cas.cn \\
liujj21@mails.tsinghua.edu.cn\\
vtakhist@post.kek.jp}

\begin{abstract}
Axion-like particles (ALPs) provide a compelling avenue for exploring physics beyond the Standard Model. In astrophysical magnetized  plasmas an ALP–photon coupling $g_{a\gamma}$ induces energy-dependent oscillations in the photon survival probability that imprint modulations on emission spectra. X-ray observations of bright spectrally-smooth sources can provide particularly sensitive probes of ultralight ALPs with masses $m_a \lesssim 10^{-11}$ eV due to long propagation distances, strong magnetic fields and high photon statistics. We present a comprehensive forecast of ALP–photon conversion in three representative systems: (i) background active galactic nuclei (AGNs) observed through foreground intracluster magnetic fields, (ii) central AGNs within their host cluster halos and (iii) Galactic X-ray binaries viewed through the Milky Way field. Using detailed simulations we assess the prospective sensitivity of high-resolution X-ray missions including XRISM, Athena and Arcus.
For typical magnetic field configurations a 320 ks XRISM observation or 8 ks Athena observation of the Perseus Cluster AGN NGC 1275 can reach down to $g_{a\gamma} \sim 5 \times 10^{-13}$ GeV$^{-1}$ at $m_a \lesssim 10^{-12}$ eV, surpassing the current best limit. We quantify the impact of magnetic field modeling, photon statistics and spectral binning strategies. Our results demonstrate the scientific potential of future X-ray observatories to probe photon–ALP coupling in previously inaccessible parameter space, offering a powerful window into physics beyond the Standard Model.
 \end{abstract}


\bigskip
 
\section{Introduction}

Axion-like particles (ALPs) can generically arise in many extensions of the Standard Model (SM), including fundamental string theories~\citep{Svrcek:2006yi,Arvanitaki:2009fg}, models with extra dimensions~\citep{Choi:2003wr}, frameworks inspired by the Peccei–Quinn mechanism~\citep{Kim:1986ax} as well as models of dark matter~\citep{Preskill:1982cy,Abbott:1982af,Dine:1982ah,Arias:2012az}.
Ultralight ALPs with masses $m_a \lesssim 10^{-11}$ eV that couple to electromagnetism via an effective two-photon vertex enable interconversion between photons and ALPs in the presence of macroscopic magnetic fields~\citep{Raffelt:1987im}. This coupling has been constrained by laboratory-based searches such as helioscopes (e.g.~CAST~\citep{CAST:2017uph}) and will be further probed by next-generation experiments like IAXO~\citep{IAXO:2019mpb}. However, large regions of the ALP parameter space, particularly at lower masses and weaker couplings, remain unexplored. Astrophysical environments with strong magnetic fields and bright photon sources provide a powerful and complementary avenue to search for ALP-induced signatures, particularly in the X-ray band.

Complementary to traditional astrophysical searches for ultralight ALPs, novel approaches are being developed to probe ALP signals across a wide range of environments and detection strategies~\citep{Antypas:2022asj}. These include multimessenger observations of ALPs from astrophysical transients, leveraging techniques such as quantum sensing~\citep{Dailey:2020sxa,Eby:2021ece,Arakawa:2023gyq,Arakawa:2024lqr,Khamis:2024oqa,Arakawa:2025hcn}. Relativistic ALPs produced in transient sources can contribute to a diffuse axion background, which can lead to observable diffuse X-ray signatures from magnetic field–induced ALP-photon conversions~\citep{Eby:2024mhd}. ALPs produced within the hot stellar cores of stars can convert to photons again through the inverse Primakoff process in the galactic magnetic fields, which will generate X-ray signals in hard band \citep{Ning:2024eky}.
These and other emerging methods offer sensitivity to distinct production mechanisms and regions of ALP parameter space, and serve as complementary probes to the high-resolution X-ray spectroscopy explored in this work.

Bright and relatively featureless X-ray spectra from astrophysical sources such as active galactic nuclei (AGNs) and Galactic X-ray binaries offer natural laboratories for probing photon–ALP mixing. As X-rays propagate over $\sim$kpc-Mpc scales through magnetized plasma, an ALP–photon coupling induces energy-dependent oscillations in the photon survival probability. These modulations manifest as narrow sinusoidal-like features imprinted on intrinsic source emission spectra. Previous searches including Chandra grating spectrometer analysis ~\citep{Reynolds:2019uqt} and CCD data analyses from Chandra and XMM–Newton~\citep{Conlon:2017qcw, Berg:2016ese}. Those works have already placed competitive constraints on ALP-photon couplings, particularly through analyses of bright targets such as the Perseus cluster AGN and stacked AGN samples. However, the limited energy resolution $\gtrsim 60$~eV of CCD detectors necessitates coarse binning, which washes out the oscillatory features and restricts sensitivity to strongest magnetic field environments. And the grating spectrometer is intrinsically weak in terms of effective area, which means that it requires more than ten times longer exposure to achieve equivalent statistics than non-dispersive detectors.

The advent of high-resolution X-ray spectrometers marks a qualitative shift in sensitivity to ALP-induced spectral signatures. The X-Ray Imaging and Spectroscopy Mission (XRISM)  launched in September 2023 equipped with Resolve, a microcalorimeter array operating over the 0.3–12~keV band, offers energy resolution of 5~eV at 6~keV~\citep{Porter2024}. Recently, it was also demonstrated that XRISM provides unique opportunities for probing heavier keV-scale decaying ALPs via line-like spectral features~\citep{Dessert:2023vyl,Zhou2024}. The upcoming Advanced Telescope for High Energy Astrophysics (Athena) mission of European Space Agency (ESA), whose new scoped concept has been recently put forth\footnote{\url{https://www.esa.int/Science_Exploration/Space_Science/NewAthena_factsheet}}, targeted achieving resolution of less than 3~eV and an effective area exceeding $1~\mathrm{m^2}$ at 1~keV~\citep{Barret:2019qaw}. Concept missions such as  Arcus aim to implement high-dispersion gratings, achieving resolving powers of $E/\Delta E > 3000$ for energy $E$ and resolution $\Delta E$ over the 12–50~\AA\ bandpass~\citep{Gunther2025}. Together, these next-generation observatories promise an order-of-magnitude improvement in both spectral resolution and photon statistics—precisely the regime where ALP-induced oscillations become resolvable for masses $m_a \lesssim 10^{-11}$~eV.  

In this work, we systematically analyze and present a comprehensive forecast of the sensitivity to ALP–photon conversion enabled by high-resolution X-ray telescopes. We simulate the instrument responses of XRISM, Athena, Arcus as well as hypothetical next-generation detectors, incorporating realistic astrophysical source models and turbulent magnetic field configurations. Our analysis focuses on three representative line-of-sight geometries: (i) a central AGN within a galaxy cluster, (ii) a background quasar observed through the outskirts of a foreground cluster, and (iii) a Galactic X-ray binary viewed through the Milky Way’s magnetic field. For each case, we compute the ALP-induced spectral modulations and perform statistical fits to evaluate the prospective sensitivity in the $(m_a, g_{a\gamma})$ parameter space. We further quantify the effects of energy resolution, photon statistics as well as spectral binning   and identify observational strategies most suited for maximizing discovery potential.

The paper is organized as follows. In Sec.~\ref{sec:alpsurvival}, we discuss ALP-photon mixing in inhomogeneous media and describe our propagation framework. In Sec.~\ref{sec:astrosystems} we introduce the astrophysical target systems we consider including central AGNs, background quasars, and Galactic X-ray sources. We also discuss details of the corresponding magnetic field and plasma models for each case, including simulations of the field introduced in Sec.~\ref{sec:simulations}. Sec.~\ref{sec:responses} summarizes our simulations of the instrument responses. In Sec.~\ref{sec:fitting} and Sec.~\ref{sec:forecast} we present our statistical analysis and resulting sensitivity forecasts for ALP-photon couplings. We conclude in Sec.~\ref{sec:conclusions}. 

\section{ALP-photon conversion}
\label{sec:alpsurvival}

A pseudoscalar ALP field $a$ can couple to photons through the interaction term in the Lagrangian
\begin{eqnarray} 
\mathcal{L}_a \supset  -\dfrac{1}{4}  g_{a \gamma} a F_{\mu\nu} \tilde{F}^{\mu\nu} = g_{a \gamma} a (\bold{E} \cdot \bold{B}) .
\end{eqnarray}
where $F_{\mu\nu}$ is the electromagnetic field strength tensor, $\tilde{F}_{\mu\nu}$ is its dual, $g_{a \gamma}$ denotes ALP-photon coupling constant  and $a$ is ALP field. Here, $\boldsymbol{E}$ and $\boldsymbol{B}$ represent the electric and magnetic fields, respectively. This coupling enables interconversion (oscillation) between photons and ALPs in the presence of an external magnetic fields~\citep{Raffelt:1987im}, such as during propagation through astrophysical environments.

Considering relativistic ALPs with $m_a \ll E$, the evolution of a photon-ALP beam with energy $E$ propagating in the $z$-direction is governed by a Schrödinger-like equation~\citep{Raffelt:1987im} 
\begin{eqnarray}
\label{equ:beam_propagation}
\centering
(i\frac{d}{dz} + E + \mathcal{M}(z) ) \begin{pmatrix}
A_x \\
A_y \\
a
\end{pmatrix} = 0~,
\end{eqnarray}
where $A_x$ and $A_y$ are the two photon polarization amplitudes, and $a$ is the ALP field amplitude. Neglecting Faraday rotation, the mixing matrix $\mathcal{M}(z)$ is given by 
\begin{eqnarray}
\centering
\mathcal{M}(z) =  \begin{pmatrix}
\Delta_{\rm pl}(z) & 0              & \Delta_{x} (z) \\
0              & \Delta_{\rm pl}(z) & \Delta_{y}(z) \\
\Delta_x (z)   & \Delta_y (z)   & \Delta_a   
\end{pmatrix}~,
\end{eqnarray}
where $\Delta_a = - m_a^2 / (2E)$, $\Delta_{\rm pl}(z) = \omega_{\rm pl}^2 / (2E)$ accounts for the effective photon mass due to the plasma, with the plasma frequency defined as $\omega_{\rm pl} = \sqrt{4 \pi \alpha n_e / m_e} \simeq 4 \times 10^{-11}~{\rm eV} \sqrt{(n_e/{\rm cm}^{-3})}$ being the plasma frequency. The off-diagonal terms describe photon-ALP mixing in the presence of a transverse magnetic field, with $\Delta_x = g_{a\gamma} B_x (z) / 2$ and $\Delta_y = g_{a\gamma} B_y (z) / 2$. 

For a beam initially in a pure polarization state $(A_x, A_y, a) = (1, 0, 0)$ propagating through a uniform magnetic field of strength $B$ aligned along the $x$-axis over a distance $L$ the photon-ALP conversion probability simplifies to
\begin{equation}
P_{A_x \rightarrow a} = \frac{\Theta^2}{1 + \Theta^2}  \sin^2\left( \Delta_{\rm eff} \sqrt{1 + \Theta^2} \right),
\end{equation}
where $\Theta = 2 B_{\perp} E g_{a\gamma}/m_{\rm eff}^2$, and $\Delta_{\rm eff} =  m_{\rm eff} L/(4E)$. The effective mass is defined as $m_{\rm eff}^2 = m_a^2 - \omega_{\rm pl}^2$, with $m_a$ denoting the ALP mass and $\omega_{\rm pl}$ the plasma frequency.
This illustrates how the photon–ALP conversion probability exhibits oscillatory behavior that depends on the photon energy, ALP mass  and the plasma properties of the medium.

To compute the photon–ALP conversion probability in the general case  we adopt the Python package ALPRO\footnote{\url{https://github.com/jhmatthews/alpro}}~\citep{Matthews2022}, which numerically solves the Schrödinger-like propagation equation for photon–ALP mixing. The code supports calculations for both pure polarization states and unpolarized photon beams.~ALPRO has been validated against multiple analytic and numerical results, including comparisons with Ref.~\citep{Marsh2017} and the GAMMAALPS~package~\citep{Meyer2014,Meyer2021}, demonstrating good consistency. The implementation includes an adaptive treatment of resonances, following the method described in~Ref.~\citep{Sisk2022}, to accurately compute ALP survival probabilities in the presence of inhomogeneous media.

\begin{figure}[t]
    \centering
    \includegraphics[width=0.5\textwidth]{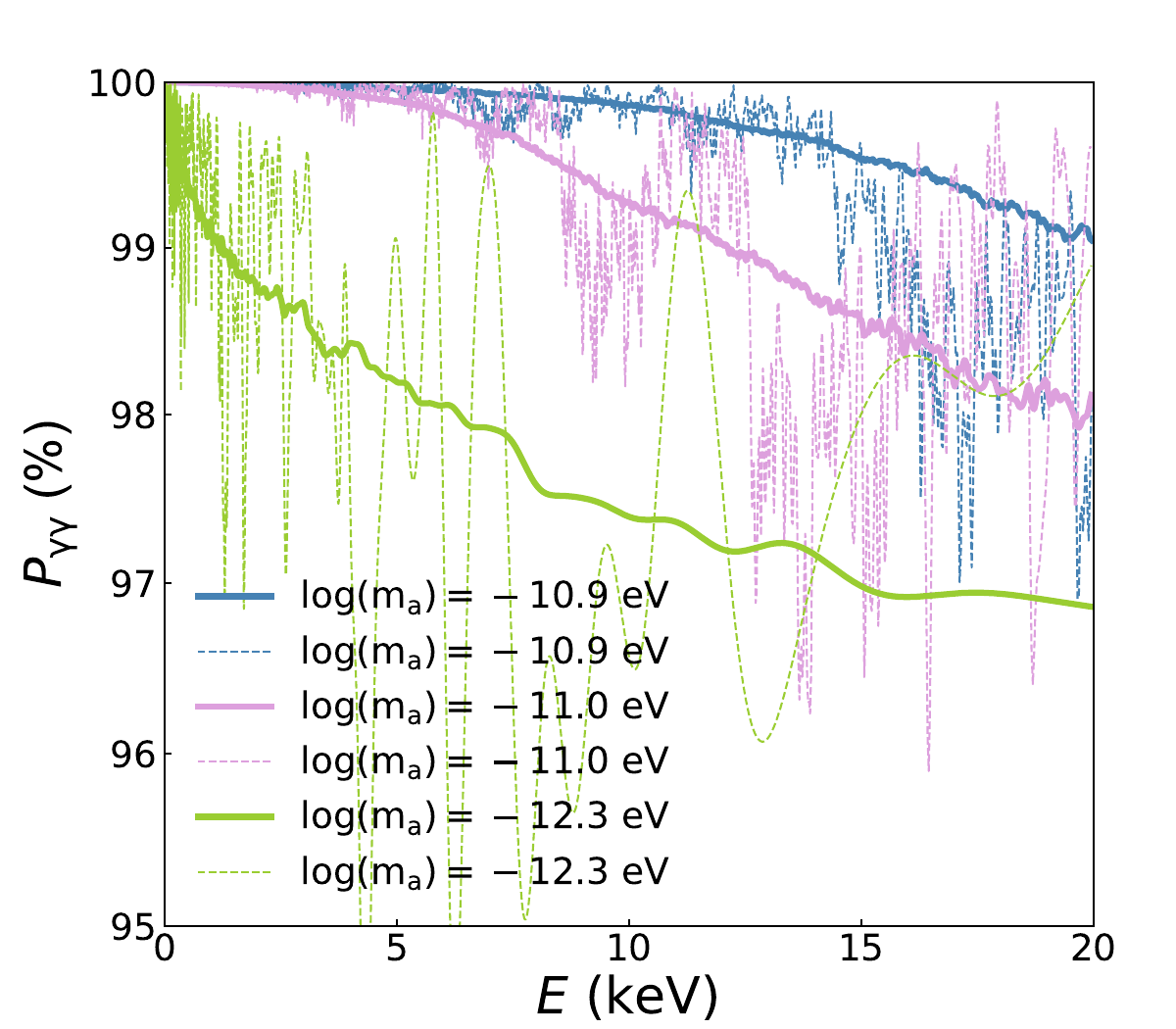}
    \caption{Photon survival probability as a function of photon energy for different ALP masses $m_a = 10^{-12.3}$ eV, $10^{-11.0}$ eV  and $10^{-10.9}$ eV considering ALP-photon coupling $g_{a\gamma} = 10^{-12}$ GeV$^{-1}$.
    The magnetic field model corresponds to the Perseus cluster as described in Sec.~\ref{section:3.1}. Solid lines show the average survival probability over 300 realizations of a randomly generated turbulent magnetic field, while dashed lines represent an example survival probability for a single random field realization.}
   \label{fig:photon_survival_probability_mass}
\end{figure}

In the astrophysical systems considered for photon–ALP beam propagation, the polarization states of the source photons are not well constrained by current observations. For example, X-ray emission from AGNs may exhibit polarization due to Compton scattering in the hot corona surrounding the accretion disk or from synchrotron processes and scattering in helical magnetic fields associated with relativistic jets~\citep{Schnittman2010, Beheshtipour2017, Titarchuk2025}. The degree and orientation of the polarization depend on the specific source scenario and viewing angle, and can range from a few percent for accretion disk coronae, i.e. $9.1\pm 1.6$ percent for black hole X-ray binary IGR J17091-3624 \citep{Ewing2025} to up to $\sim 30\%$ for jet-dominated emission in blazar PKS 2155-304 as observed by IXPE~\citep{Kouch2024,McNamara2009,Titarchuk2025}.

Given the lack of detailed measurements of the X-ray polarization for these sources, we do not assume any specific initial polarization state in our calculations. Instead, we compute the unpolarized photon survival probability by averaging over two orthogonal pure polarization states following 
\begin{equation}
P_{\gamma\gamma} = 1 - P_{\gamma a} = 1 - \frac{1}{2}(P_{A_x \rightarrow a} + P_{A_y \rightarrow a})~,
\end{equation}
where $P_{\gamma a}$ is the total unpolarized photon-ALP conversion probability. 

In Fig.~\ref{fig:photon_survival_probability_mass} we show the photon survival probability as a function of energy, computed using the magnetic field and electron density profiles of the Perseus cluster, as described below. The photon–ALP conversion induces rapid energy-dependent oscillations in the survival probability resulting in characteristic absorption-like features. The width of these spectral modulations can be as narrow as sub-keV above 5~keV and less than 200~eV below 3~keV. While such fine structures are largely smeared out in CCD detectors due to their limited energy resolution, they remain detectable with cryogenic microcalorimeters, such as XRISM Resolve that achieves an energy resolution of approximately 5~eV.

 \begin{figure*}[t]
    \centering
    \includegraphics[width=0.7\textwidth]{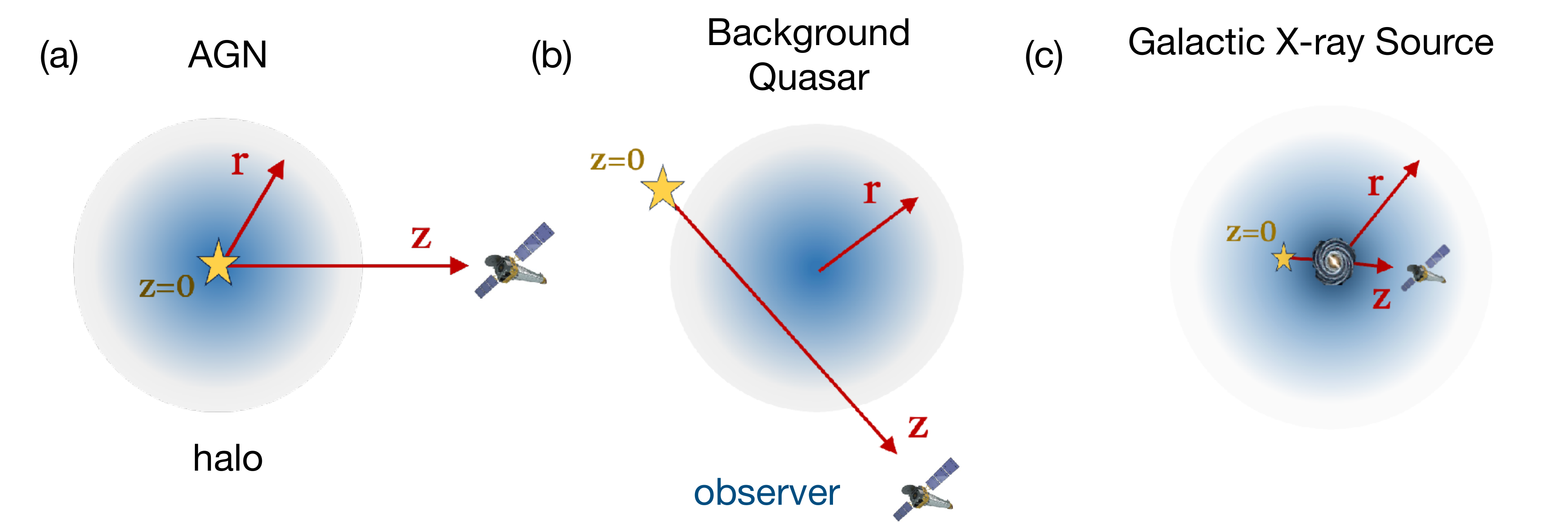}
    \caption{Illustration of the three astrophysical systems considered in this study: (a) AGN  located at the center of a galaxy cluster,
(b) background quasar with its sight line passing through the outskirts of a galaxy cluster halo  and
(c) X-ray binary situated in the Milky Way's Galactic halo.
The yellow star indicates the position of the photon source. The blue z-axis line represents the photon beam's propagation direction, along which conversion into ALPs occurs. The point $z=0$ marks the start of the photon-ALP conversion region.
}
    \label{fig:target}
\end{figure*}

\section{Astrophysical Target Systems}
\label{sec:astrosystems}

Various astrophysical systems provide suitable conditions for investigating photon–ALP conversion, where a bright background photon source propagates through a macroscopic magnetic field along the line of sight. In this work, we investigate three representative classes of such systems and select optimal targets for each, to demonstrate their capability in constraining photon–ALP mixing. The target categories we consider are: (i)  AGN  located at the centers of galaxy clusters, (ii) background quasars with sight lines passing through the outskirts of a foreground galaxy cluster halo  and (iii) Galactic X-ray binaries, whose emission traverses the Milky Way's magnetic field.

Fig.~\ref{fig:target} illustrates schematically the geometric configurations of the three systems we consider. In Tab.~\ref{tab:summary of targets} we summarize the basic properties of the selected target sources and their associated magnetic field environments.  In particular,
NGC 1275 located in the Perseus cluster is the brightest known AGN associated with a galaxy cluster, making it an ideal target for probing photon–ALP conversion. However, constraints derived from a single system like NGC 1275 may be subject to systematic uncertainties arising from the stochastic nature of AGN activity and the turbulent intracluster magnetic field. To mitigate this, we conducted an extensive search for background quasars whose sight lines partially intersect the magnetic fields of galaxy clusters with existing measurements of field properties. Among all identified sources, we select the background quasar with the highest flux, SDSS J162904.36+393417.7, behind the galaxy cluster Abell 2199, as the target. In addition, we consider Galactic X-ray sources as potential targets. For these systems, the photon beam propagates through the Milky Way's magnetic field. Since Galactic sources are located at much closer distances compared to extragalactic targets, it is important to assess the competing effects of higher photon statistics versus the shorter propagation path on the overall sensitivity to photon–ALP conversion. We optimize the choice by taking both source brightness and magnetic field strength into consideration, and select the X-ray binary source 4U 1820-30 as the representative target for the study of Galactic system. 

\begin{figure*}[t]
    \centering
    \includegraphics[width=1\textwidth]{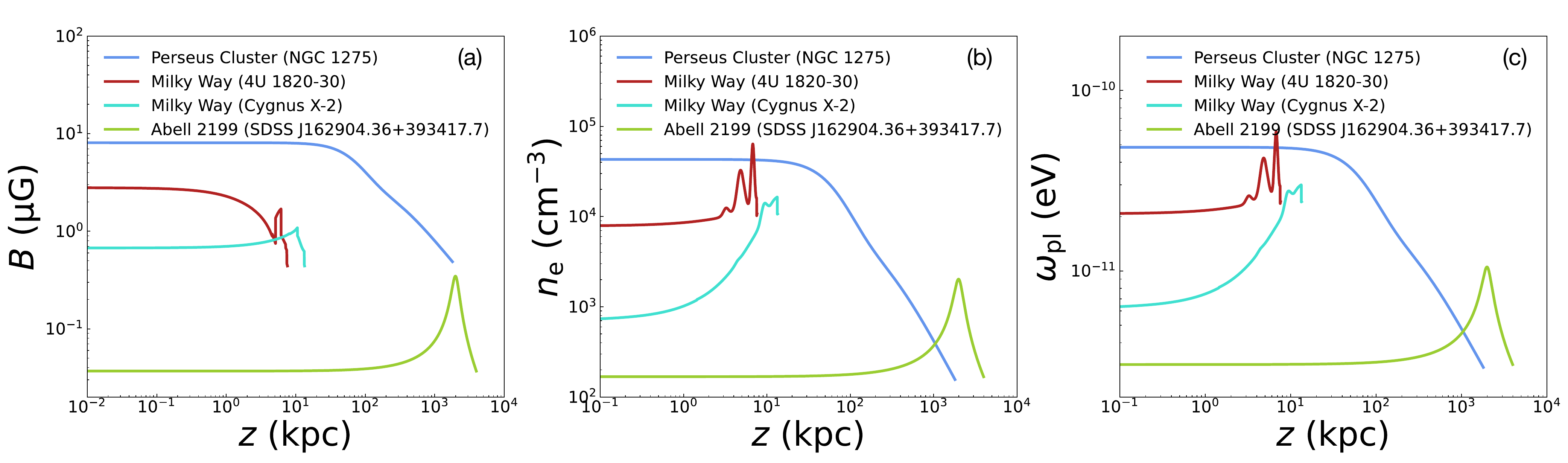}
    \caption{Profile distributions of (a) magnetic flux density, (b) electron density and (c) plasma frequency of the intracluster or interstellar medium along the photon beam propagation path in three astrophysical systems we consider: the Perseus cluster, the Milky Way and Abell 2199. The $z$-axis is defined as the distance from the background photon source, with $z=0$ corresponding to the source location as illustrated in Fig.~\ref{fig:target}.}
    \label{fig:Eletron-density & Magnetic field for galaxy clusters & Milky Way.}
\end{figure*}

Fig.~\ref{fig:Eletron-density & Magnetic field for galaxy clusters & Milky Way.} shows the magnetic field, electron density, and plasma frequency of the three astrophysical target systems we consider.

\begin{deluxetable*}{lcccc}[t]
\label{tab:summary of targets}
\tablenum{1}
\tablewidth{0pt}
\tabletypesize{\normalsize}
\tablecaption{Basic information of the three astrophysical target systems considered: (a) central AGN in a galaxy cluster, (b) background quasar behind a foreground cluster  and (c) Galactic X-ray binary. Astrophysical source environment through which the X-ray signal propagates along the line of sight is indicated. }
\tablehead{
\colhead{Type} & \colhead{Source} & \colhead{Flux in 2--10 keV} & \colhead{Magnetic Field} & \colhead{Line of Sight} \\
\colhead{} & \colhead{} & \colhead{(ergs cm$^{-2}$ s$^{-1}$)} & \colhead{Environment} & \colhead{$|B_{\rm max}|$ ($\mu$G)}
}
\startdata
\multirow{1}{*}{(a) Central AGN} & NGC 1275 & $2.209 \times 10^{-9}$ & Perseus Cluster & 8 \\
\multirow{1}{*}{(b) Background quasar} & SDSS J162904.36+393417.7 & $1.0745 \times 10^{-13}$ & Abell 2199 Cluster & 0.4 \\
 \multirow{1}{*}{(c) Galactic X-ray binary} & 4U 1820--30 & $5.0048 \times 10^{-9}$ & Milky Way & 3 \\
\enddata 
\end{deluxetable*}

\subsection{Central AGN in Galaxy Cluster}
\label{section:3.1}

The X-ray emission of AGNs is typically characterized by a featureless continuum, making these sources ideal backgrounds for searches of photon–ALP conversion. Since this phenomenon requires a magnetized environment extending over large volumes, galaxy clusters offer particularly favorable conditions. Their intracluster medium hosts magnetic fields with typical strengths of several $\mu$G, coherently extending over megaparsec scales providing a natural laboratory for photon–ALP oscillations. Among known systems NGC 1275 located at the center of the Perseus cluster is the brightest AGN associated with a galaxy cluster~\citep{Snowden2008}. This makes NGC 1275 the optimal target for our study. 

AGN X-ray emission originates primarily from accretion disk photons that are Compton up-scattered by a surrounding corona of hot electrons. This process typically produces a power-law energy spectrum, that is, the photon energy distribution following the form $\propto E^{-\Gamma}$, usually with a photon index $0< \Gamma < 2$. The observed emission is further modified by photoelectric absorption, which can occur in the obscuring dusty torus surrounding the AGN within the intracluster medium of the host galaxy cluster or in the interstellar Galactic medium of the Milky Way along the line of sight.

For our simulations we phenomenologically model AGN X-ray spectra as power-law using \texttt{TBabs*power-law} in \texttt{XSPEC} package\footnote{https://heasarc.gsfc.nasa.gov/docs/xanadu/xspec/}, where \texttt{TBabs} represents the photoelectric absorption component. For NGC 1275, we adopt model parameters consistent with fits to the Chandra grating spectrum with 490 ks of on-source exposure \citep{Reynolds:2019uqt}. Specifically, we use a hydrogen column density of $n_{\rm H} = 1.32 \times 10^{21}~ \mathrm{cm}^{-2}$, a photon index of $\Gamma = 1.88$, and a normalization of $A = 8.34 \times 10^{-3}$, corresponding to a flux of $3.18 \times 10^{-11}~ \mathrm{ergs~cm^{-2} ~ s^{-1}}$ in the 1–9 keV band.

Although more complex spectral models exist to account for reprocessed emission in Compton-thin plasma—such as the iron K$\alpha$ line at 6.4 keV or partial covering absorption from clumpy molecular gas in the cluster core \citep{Matthews2022}—these are beyond the scope of this study. Since our goal is not to fit observational data but rather to demonstrate the sensitivity reach of current and future X-ray detectors to photon–ALP conversion, we adopt the absorbed power-law model as the input for simulating the target source spectrum.

We model the propagation medium for photons and their conversion into ALPs by parameterizing the magnetic field strength as a function of distance $z$ from the cluster center. The field is assumed to follow a density dependent power-law 
\begin{eqnarray}
\label{equ : B-field}
\centering
B(z) = B_0 \Big[\frac{n_e (z)}{n(R_0)}\Big]^{\alpha}~,
\end{eqnarray}
where $R_0 = 25~\mathrm{kpc}$ is a reference radius, $B_0 = 7.5~  \mu\mathrm{G}$ is the magnetic field normalization  and index $\alpha = 0.5$. These parameters are based on the radial pressure profile of the Perseus cluster derived by Ref.~\citep{Fabian2006} from deep X-ray observations.

For the electron density profile of the Perseus cluster we adopt the double-$\beta$ model of Ref.~\citep{Churazov2003}. Namely,
\begin{align} 
n_e (z) =&~ \frac{3.9\times 10^{-2}}{ [1 + (z/80\ {\rm kpc})^2 ]^{1.8}} {\rm cm}^{-3} \notag\\ &+ \frac{4.05\times 10^{-3}}{[1+ (z/280\ {\rm kpc})]^{0.87}} {\rm cm}^{-3}~.
\end{align}

\subsection{Background Quasar Behind Galaxy Cluster}
\label{section:QSO-cluster}

In addition to central AGNs residing in galaxy clusters, systems where a background quasar has a line of sight that partially intersects the outskirts of a foreground galaxy cluster also provide suitable environments for studying photon–ALP conversion. In these configurations, the X-ray photon beam from the quasar propagates through the cluster's magnetized intracluster medium, offering another opportunity to probe ALP-induced effects.
To identify favorable quasar–galaxy cluster pairs, we compiled a list of radio-loud galaxy clusters with well-characterized magnetic fields  based on measurements from diffuse synchrotron emission and Faraday rotation studies~\citep{Vacca2018}. The magnetic field power spectra in these clusters have been extensively modeled through numerical simulations to reproduce the observed Faraday rotation measures along known lines of sight.

We cross-correlated the coordinates of these galaxy clusters with X-ray background quasar catalogs from \texttt{SIMBAD}\footnote{https://simbad.cds.unistra.fr/simbad/} and identified quasars whose projected locations lie near the outskirts of the selected clusters. The relevant properties of the selected galaxy clusters and their corresponding background quasars are summarized in Tab.~\ref{tab:qso_info}.
For these sources, we analyzed all available XMM-Newton observations of the selected background quasars and extracted their combined X-ray spectra. The data reduction procedures are described below. Among the identified candidates, we selected the brightest quasar as the primary target for further analysis. 

We carry out the spectrum fitting with \texttt{XSPEC} for all selected background quasars and the observed fluxes in 0.3–10.0 keV are summarized in Tab.~\ref{tab:qso_info} in App.~\ref{app:secqso}. The details of data reduction are described in App.~\ref{app:secqso}. Most sources are limited by low photon statistics so that only a simple model \texttt{phabs * power-law} is used for spectrum fitting. In the case of UGC 12064, where the spectrum shows optically thin plasma emission in the 0.5–1.0 keV range, we adopt a more complex model given by \texttt{phabs*zTBabs*(apec+zpower-law)}. Here, \texttt{phabs} accounts for the Galactic absorption, fixed to the hydrogen column density from the HI4PI survey \citep{HI4PI2016}, while the \texttt{zTBabs} component models the absorption in both the foreground cluster and the quasar host galaxy. 

Among all identified quasar–cluster pairs, we select the source with the highest observed flux in 0.3–10 keV band, SDSS J162904.36+393417.7, as our best target for the ALP-photon sensitivity demonstration study. The spectrum is modeled as \texttt{TBabs*power-law} in \texttt{XSPEC}, with a column density of $n_{\rm H} = 5.39 \times 10^{20}~\mathrm{cm^{-2}}$ and a photon index of $\Gamma = 1.80$.

We neglect the contribution from the quasar host galaxy's magnetic field in this analysis. Observations of galaxies from $z=0\sim1$ indicate no significant evolution in large-scale magnetic field strengths over cosmic time~\citep{Bernet_2008_magneticfield_z_evolution}. Local galaxies typically exhibit fields of order $\sim3~\mu\mathrm{G}$ \citep{Han_2006_magneticfield}, which is subdominant compared to the cluster field modeled here.
The effect of the intergalactic magnetic field is also neglected, as current upper limits suggest typical field strengths of $\lesssim~ \mathrm{nG}$~\citep{Pomakov_2022_IGM_Bfield}, orders of magnitude below the cluster magnetic field and thus irrelevant for photon–ALP conversion in this context.
We model the magnetic field profile of Abell 2199 using Eq.~\eqref{equ : B-field} considering input parameters
$B_0 = 11.7~\mu$G, $\alpha = 0.9$ and $n_0 = 0.101~{\rm  cm}^{-3}$.

The electron density of Abell 2199 is parameterized using a double-$\beta$ model 
\begin{equation}
\label{equ:A2199 electron-density profile} 
n_e =~ n_{0_{i}}  \Big(1 + \dfrac{r^2}{r_{c_{i}}^2} \Big)^{-1.5 \beta_{i}}  
 + n_{0_{e}} \Big(1 + \dfrac{r^2}{r_{c_{e}}^2} \Big)^{-1.5 \beta_{e}}~,
\end{equation}
with parameters $n_{0_{i}} = 0.074\rm\ cm^{-3}$, $r_{c_{i}} = 0.009\rm\ Mpc$, $\beta_{i} = 1.5$, $n_{0_{e}} = 0.027\rm\ cm^{-3}$, $r_{c_{e}} = 0.026\rm\ Mpc$, $\beta_{e} = 0.39$.

\subsection{Galactic X-ray Sources in Milky Way}
\label{section:MW}

The Milky Way hosts Galactic magnetic field that has been extensively modeled and constrained through multiple datasets, including Faraday rotation measures, synchrotron emission, and polarized starlight~\citep{Cordes2002,Jansson2012,Pshirkov2011}. In addition, the Galaxy contains several black hole low-mass X-ray binary (LMXB) systems, some of which exhibit simple X-ray spectra that make them suitable as background photon sources for photon–ALP conversion studies. These sources are typically nearby, offering high fluxes and favorable signal statistics. However, their proximity also limits the available propagation distance for ALP conversion, reducing the overall sensitivity compared to extragalactic cases.

We identify the most suitable Galactic target by  systematically analyzing LMXB systems from the catalog of Ref.~\citep{Luo2018}. Extremely bright sources such as Sco X-1 were excluded due to possibility of detector saturation. After consideration of flux, distance and spectral simplicity, we identified 4U 1820–30 as an optimal Galactic source for photon–ALP conversion studies. This system has a high X-ray flux and is located at a distance of $7.6 \pm 0.4~ \mathrm{kpc}$ from Earth. Furthermore, its low Galactic latitude, positioning it near the Galactic plane, ensures that the photon beam propagates through regions of the Milky Way with relatively strong magnetic fields  enhancing the potential for ALP-induced spectral effects.

The X-ray energy spectrum of 4U 1820–30 is well described by an absorbed blackbody plus Compton emission model, represented in \texttt{XSPEC} as \texttt{TBabs*(bbody+compTT)}. In this model, the galactic neutral and ionized interstellar medium absorption is modeled with an equivalent hydrogen column density of $n_{\rm H} = 7.8 \times 10^{20}~ \mathrm{cm}^{-2}$. The blackbody component has a temperature of $kT_{\rm bb} = 0.26~ \mathrm{keV}$, while the Comptonization component is characterized by a seed photon temperature of $kT_0 = 0.48~ \mathrm{keV}$, an electron temperature of $kT_1 = 3.9~ \mathrm{keV}$  and an optical depth of $\tau = 5.4$. 
We assume the source is in a relatively high flux state, consistent with previous observations. The 2–10 keV flux is set to $5.02 \times 10^{-9}~ \mathrm{ergs~ cm^{-2}~ s^{-1}}$, based on INTEGRAL JEM-X measurements, while the 0.5–2 keV flux is $1.32 \times 10^{-9}~ \mathrm{ergs~ cm^{-2}~ s^{-1}}$, obtained from XMM-Newton RGS data~\citep{Costantini2012}.

The Galactic electron density profile along the line of sight is computed using the YMW16 model~\citep{Yao2017}, which fits dispersion measures of 189 pulsars with independently measured distances. We employ the~\texttt{pygedm} Python package\footnote{\url{https://github.com/FRBs/pygedm}} to evaluate the electron density for the specific line of sight corresponding to 4U 1820–30. 

The Galactic magnetic field is modeled using the framework of Ref.~\citep{Jansson2012}, which combines fits to the WMAP7 Galactic synchrotron emission map and over 40,000 extragalactic rotation measures. This model incorporates contributions from the disk, extended halo  and out-of-plane field components. We compute the magnetic field amplitude along the selected line of sight using the \texttt{gmf} package\footnote{\url{https://github.com/me-manu/gmf}}, implemented with a modified version of the NE2001 code \citep{Cordes2002} to ensure consistency with electron density modeling.

\section{Simulations of Turbulent Magnetic Field}
\label{sec:simulations}

The magnetic field in galaxy clusters is expected to be turbulent on $\sim$kpc scales, as indicated by both observations \citep{Zhuravleva2014} and simulations \citep{Donnert2018}. Numerically, magnetic turbulence can be modeled using various approaches that randomize the field structure at specific coherence lengths. Since the exact magnetic field configuration is unknown and expected to be turbulent, we compute the photon survival probability over an ensemble of model realizations and average the final ALP-photon survival probability results. This reflects the effective behavior expected in real data and avoids over-interpreting fluctuations tied to any particular field configuration.

\subsection{Cell-based Model}
\label{sec:cell-based}

A commonly adopted method to simulate astrophysical turbulent magnetic fields is the cell-based model, where the magnetic field is represented as a series of ``cell'' regions of size $\Delta z$, corresponding approximately to the local coherence length $\Lambda_c$. Each cell maintains a constant magnetic field direction and strength, but these properties change randomly between adjacent cells.
The sizes of the cells are drawn from a power-law probability distribution
\begin{equation}
p(\Delta z) \propto \Delta z^{-n}~,
\end{equation}
where $n$ is the spectral index of the turbulence, and cell sizes span a specified range of $[\Lambda_{\rm min},~ \Lambda_{\rm max}]$. This allows the model to capture the multi-scale nature of magnetic turbulence.

To account for the expected radial variation of turbulence in galaxy clusters the minimum and maximum cell sizes can be scaled linearly with radius. This reflects the tendency for coherence lengths to increase with distance from the cluster center.

\subsection{Gaussian Random Field}
\label{sec:GRF}

We also simulate the turbulent magnetic field using an alternative Gaussian Random Field (GRF) approach. Here, the spatial cell resolution is fixed at a grid size $\delta z$ and the field is characterized by an isotropic power spectrum of the form $E_k dk \propto k^{-n} dk$, where $E_k\ dk$ is the magnetic energy contained in the wavenumber range $(k, k+dk)$ with $k = 2\pi / \Lambda$ representing the wavenumber corresponding to spatial scale $\Lambda$. The spectral index $n$ is set to $5/3$, considering Kolmogorov turbulence.

The coherence (correlation) length of the turbulent field is then given by
\begin{eqnarray}
\centering
\Lambda_c = \frac{\Lambda_{\rm max}}{2}  \frac{n-1}{n} \frac{1 - (\Lambda_{\rm max} / \Lambda_{\rm min})^n }{1 - (\Lambda_{\rm max} / \Lambda_{\rm min})^{n-1}},
\end{eqnarray}
where $\Lambda_{\rm min}$ and $\Lambda_{\rm max}$ denote the minimum and maximum turbulence scales, set to 3.5 kpc and 30 kpc, respectively.

To generate GRF, we first compute random Fourier amplitudes following a Rayleigh distribution
\begin{equation}
P(A)dA = \frac{A}{2\pi A_k^2} \exp\left(-\frac{A^2}{A_k^2}\right) dA~,
\end{equation}
where $A_k \propto k^{-\eta}$ is the amplitude scaling with wavenumber, and $\eta = n + 2$ to account for the 3-dimensional field construction. The wavenumbers $k$ range between $2\pi/\Lambda_{\rm max}$ and $2\pi/\Lambda_{\rm min}$.

The Fourier phases $\phi$ are drawn from a uniform distribution between $0$ and $2\pi$. We generate the vector potential $\vec{A}(\vec{k})$ in 3-dimensions, then compute the magnetic field in Fourier space using
\begin{equation}
\vec{B}(\vec{k}) = i \vec{k} \times \vec{A}(\vec{k})~.
\end{equation}
We then perform an inverse Fourier transform in order to obtain the magnetic field $\vec{B}(\vec{r})$ in real space.

The normalization constant $C$ is chosen such that the average magnetic field amplitude within the cluster core radius $r_c$ matches the central field strength $B_0$. This is computed as
\begin{equation}
C = \dfrac{N_{r<r_c}}{ \sum_{r<r_c} \left[ B_{\rm amp}(r)\left( \dfrac{n_e(r)}{n_e(0)} \right)^{\alpha} \right]},
\end{equation}
where $N_{r<r_c}$ is the number of $ B_{\rm amp}$ grid points within the core radius with $r < r_c$. And the denominator computes the summation of the magnetic field within $r_c$, where $B_{\rm amp}$ is the non-normalized generated random field amplitude and the factor $\left( n_e(r)/n_e(0) \right)^{\alpha}$ incorporates the radial scaling of the magnetic field with the cluster electron density profile, following Ref.~\citep{Angus2014}. After normalization, the perpendicular component of the magnetic field is adjusted to match the specified cluster $B$-field profile.

\begin{figure}[t]
    \centering
    \includegraphics[width=0.5\textwidth]{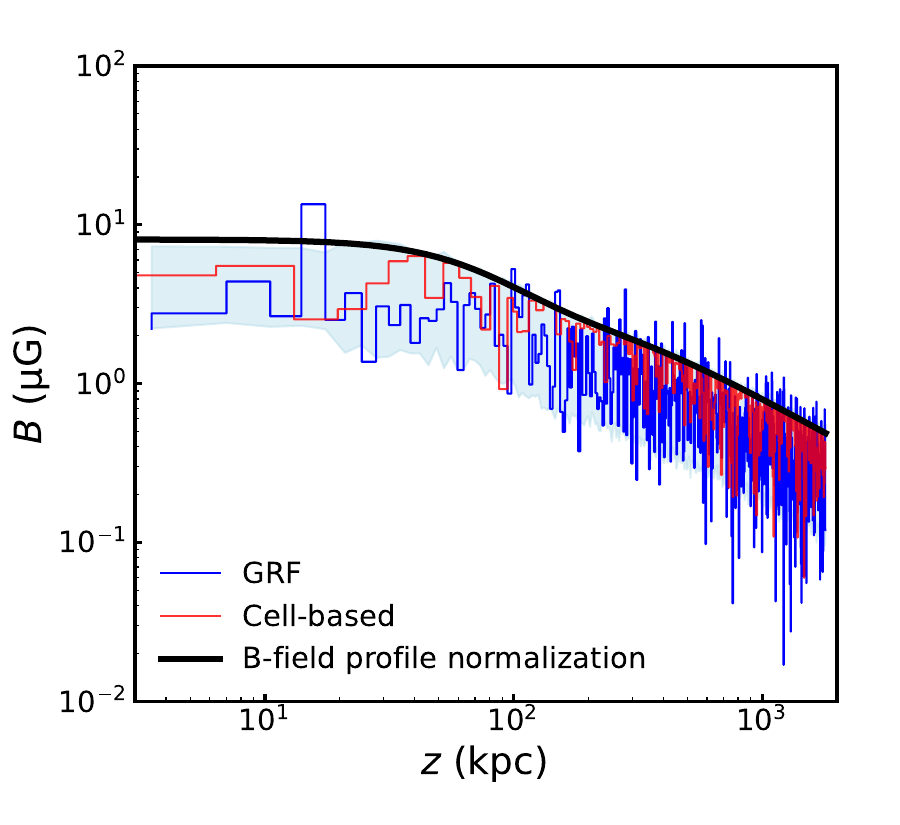}
    \caption{Radial profiles of the turbulent magnetic field models generated using the cell-based algorithm (red) and GRF method (blue). The black line indicates the underlying regular magnetic field profile defined by Eq.~\eqref{equ : B-field} that serves as the normalization reference for the turbulent field models. The blue shaded region represents the standard deviation of the GRF realizations computed over 400 simulations.}
    \label{fig:Bfield_model}
\end{figure}

In  Fig.~\ref{fig:Bfield_model} we show the radial profiles of the turbulent magnetic field realizations generated using both the cell-based algorithm and GRF method, compared to the underlying regular magnetic field profile.
Although both turbulence models reproduce the average field strength profile by construction the differences in their stochastic realizations could introduce systematic variations in the final ALP sensitivity forecasts. To quantify this, we simulate the ALP-photon conversion probability using both methods and compare the results in App.~\ref{section:compare-B-field-models}.

\subsection{Simulations for Perseus Cluster}
\label{sec:B-field_perseus}

To account for the turbulent nature of the intracluster magnetic field in Perseus cluster, we employ a cell-based algorithm that generates a random field configuration with a  coherence length distribution following $p(\Delta z) \propto \Delta z^{-1.2}$ ranging from 3.5 kpc to 10 kpc, consistent with model B in Ref.~\citep{Reynolds:2019uqt} that matches Faraday rotation observations of \citet{Taylor2006}. The cell sizes are scaled linearly with radius as $\Delta z \propto z/50\ \mathrm{kpc}$, reflecting the expected growth of magnetic coherence lengths with increasing distance from the cluster center. The radial normalization of the magnetic field is set by the profile in Eq.~\eqref{equ : B-field}.

We generate 400 independent realizations of the turbulent field as the magnetic field model, which is used in the ALP-photon conversion calculations, and compute the ensemble-averaged photon survival probability curve. For comparison, we also construct a GRF model with a Kolmogorov turbulence power spectrum and generate 400 additional realizations to obtain the corresponding average field profile.

\subsection{Simulations for Abell 2199}
\label{sec:B-field_A2199}

The photon beam from SDSS J162904.36+393417.7 passes through the outskirts of the galaxy cluster Abell 2199, defining the photon–ALP conversion region in our model. We simulate the cluster magnetic field using a cell-based turbulent field algorithm, where the cell size distribution follows $p(\Delta z) \propto \Delta z^{-2.8}$ over scales of 0.7–35 kpc. This corresponds to the minimum and maximum fluctuation scales of the magnetic field, consistent with a power-law power spectrum of index $n=2.8$~\citep{Vacca2012}. The radial profile of the magnetic field and electron density in Abell 2199 is determined as the description in Sec.~\ref{section:QSO-cluster}. 400 realizations of the magnetic field was generated as well as the ALP-photon conversion probability curves, from which we compute the ensemble-averaged photon survival probability curve that is later used in the spectrum fitting.

\subsection{Simulations for Milky Way}
\label{sec:B-field_perseus_milkyway}

Faraday rotation measure from extragalactic pulsars have revealed large-scale patterns inside the Milky Way that are signatures of coherent fields generated by dynamos \citep{Beck2001,Beck2008,Beck2009}. Further, additional features of magnetic field reversals at kpc-scales have been resolved, such as the one inside the solar radius near the southern Galactic plane~\citep{Brown2007}  and those near the Sagittarius-Carina spiral arm seen from the northern Galactic pole~\citep{Frick2001, Han2002, Han2006, Rand1994}.

For our simulations of magnetic field in Milky Way we assume regular large scale field model as described in Sec.~\ref{section:MW}, and consider  a constant coherence length of 1 kpc throughout without any coherence length scaling. We generate 400 realizations of random magnetic field for each assumed ALP mass and obtain the photon survival probability. We then compute the ensemble-average for spectrum fitting.

\section{Detector Response Modeling}
\label{sec:responses}

\begin{figure}[t]
    \centering
    \includegraphics[width=0.5\textwidth]{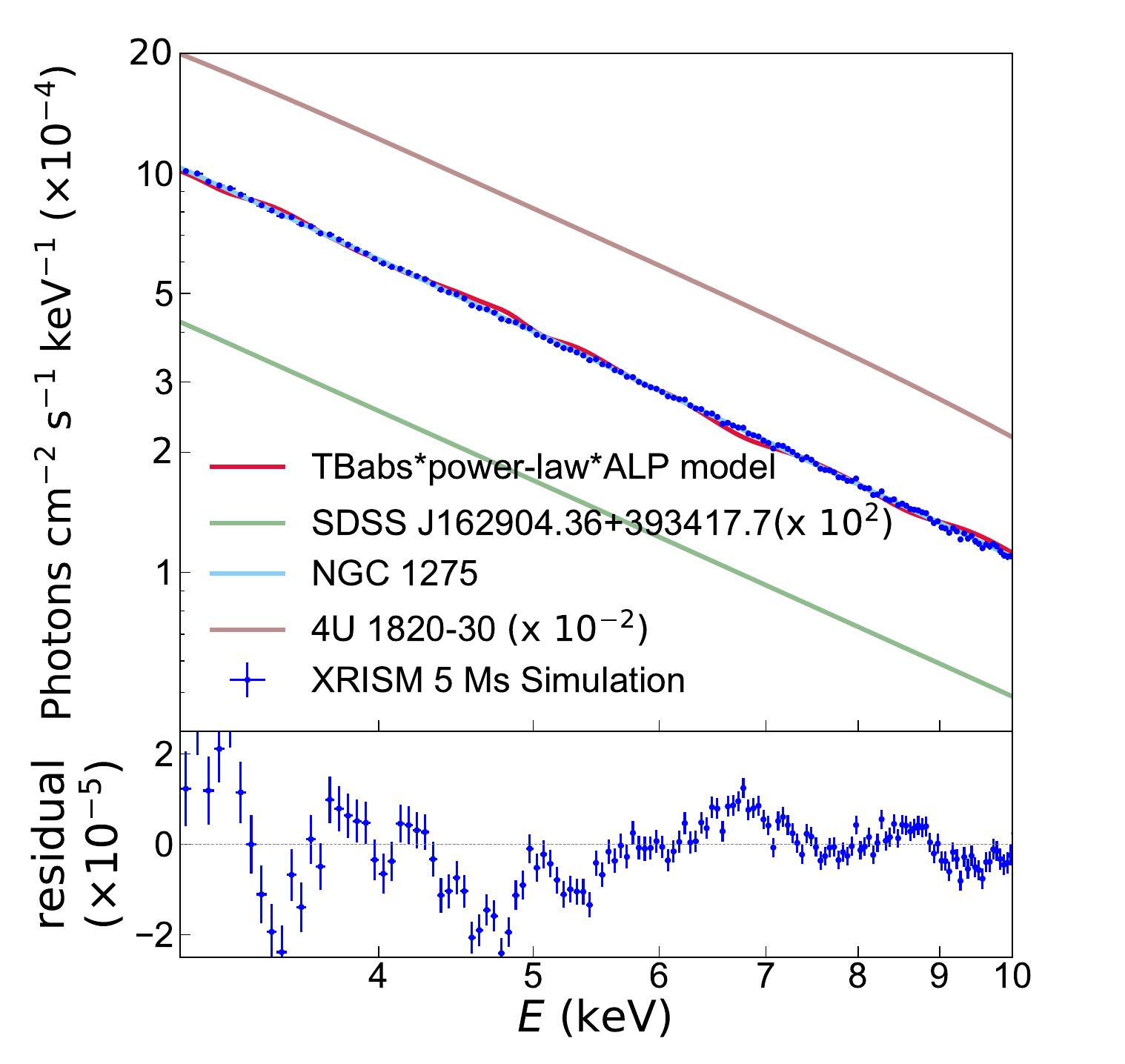}
    \caption{Simulated energy spectra of three astrophysical sources considered in this study, NGC 1275 (light blue), quasar SDSS J162904.36+393417.7 (green, scaled up by a factor of 10$^2$) and Galactic X-ray binary 4U 1820–30 (brown, scaled down by a factor of $10^{-2}$). The blue data points with error bars show the simulated spectrum assuming a 5~Ms XRISM observation. The residual panel displays the difference between the simulated data and the best-fit \texttt{ALPabs*power-law} model, assuming ALP parameters of $m_a = 5\times10^{-13}\ \mathrm{eV}$ and $g_{a\gamma} = 10^{-12}\ \mathrm{GeV}^{-1}$, with a photon index of $\Gamma = 1.88$. }
    \label{fig:energy_spec_comparison}
\end{figure}

\subsection{XRISM}

XRISM's high-resolution spectrometer Resolve enables precise X-ray spectroscopy with unprecedented spectral sensitivity. Fig.~\ref{fig:effective area} shows XRISM's effective area as a function of photon energy, comparing conditions with the gate valve open and closed. Here we perform simulations using XRISM response files considering that the gate valve is closed, which corresponds to the current operational configuration in orbit \citep{Tsujimoto2018}. The condition of closed gate valve compared to open gate valve mostly deteriorates sensitivity below 2 keV and has only mild impact on the effective area above 2 keV. Given the ALP mass and photon energy we consider in this work, the ALP-photon conversion driven oscillation feature in the background energy spectrum is more dominant in the hard energy band above 2 keV. Therefore we note that the gate valve condition will not affect significantly to our conclusion, and we adopt the response files as the gate valve closed in our simulation.

We simulate the observed energy spectrum of the target sources using XRISM Resolve, assuming on-axis pointing and exposure times of 50 ks, 200 ks, 500 ks, and 5 Ms. The simulations use the following response and background files\footnote{\url{https://xrism.isas.jaxa.jp/research/proposer/obsplan/response/index.html}}. The effective area curve used in our simulations is shown in Fig.~\ref{fig:effective area}.

Fig.~\ref{fig:energy_spec_comparison} shows the simulated XRISM spectrum of NGC 1275, assuming a total exposure of 5 Ms, demonstrating XRISM’s capability to resolve subtle features in the energy range relevant for ALP searches. 

\subsection{Athena}
Athena  is a next-generation large class X-ray observatory under development by the ESA as part of its Cosmic Vision program~\citep{Barret:2019qaw}. Athena will employ a cutting-edge transition-edge sensor array as its X-ray Integral Field Unit (X-IFU), enabling high-resolution spectroscopy across a wide field of view. The X-IFU instrument is designed to deliver an unprecedented energy resolution of approximately 2.5 eV at 7 keV, surpassing the capabilities of current X-ray missions. This will enable detailed spectroscopic studies of hot plasma, AGN and compact astrophysical objects\footnote{\url{https://apc.u-paris.fr/FACe/en/athena}}.

To simulate the instrument’s response to various source models and to evaluate its sensitivity to ALP-induced spectral features across the relevant energy range we employ Athena X-IFU response files\footnote{\url{https://x-ifu.irap.omp.eu/en/resources/for-the-community}}.
The effective area curve used in our simulations is shown in Fig.~\ref{fig:effective area}.

\subsection{Arcus}

The Arcus observatory is a proposed medium-class explorer mission designed to carry a high-resolution X-ray grating spectrometer (XGS) alongside a medium-resolution far-ultraviolet spectrometer. The X-ray instrument combines silicon pore optics with high throughput critical-angle transmission gratings, achieving a resolving power of $E/\Delta E > 3000$ over the 12–50~\AA\ bandpass~\citep{Gunther2025}. With an effective area exceeding $250~\mathrm{cm^2}$ at 19~\AA, Arcus would offer a significant advancement in soft X-ray spectroscopy. This enables precise measurements of narrow spectral features with sensitivity beyond that of current instruments ~\citep{Smith2021SPIE}.

\begin{figure}
    \centering
    \includegraphics[width=0.5\textwidth]{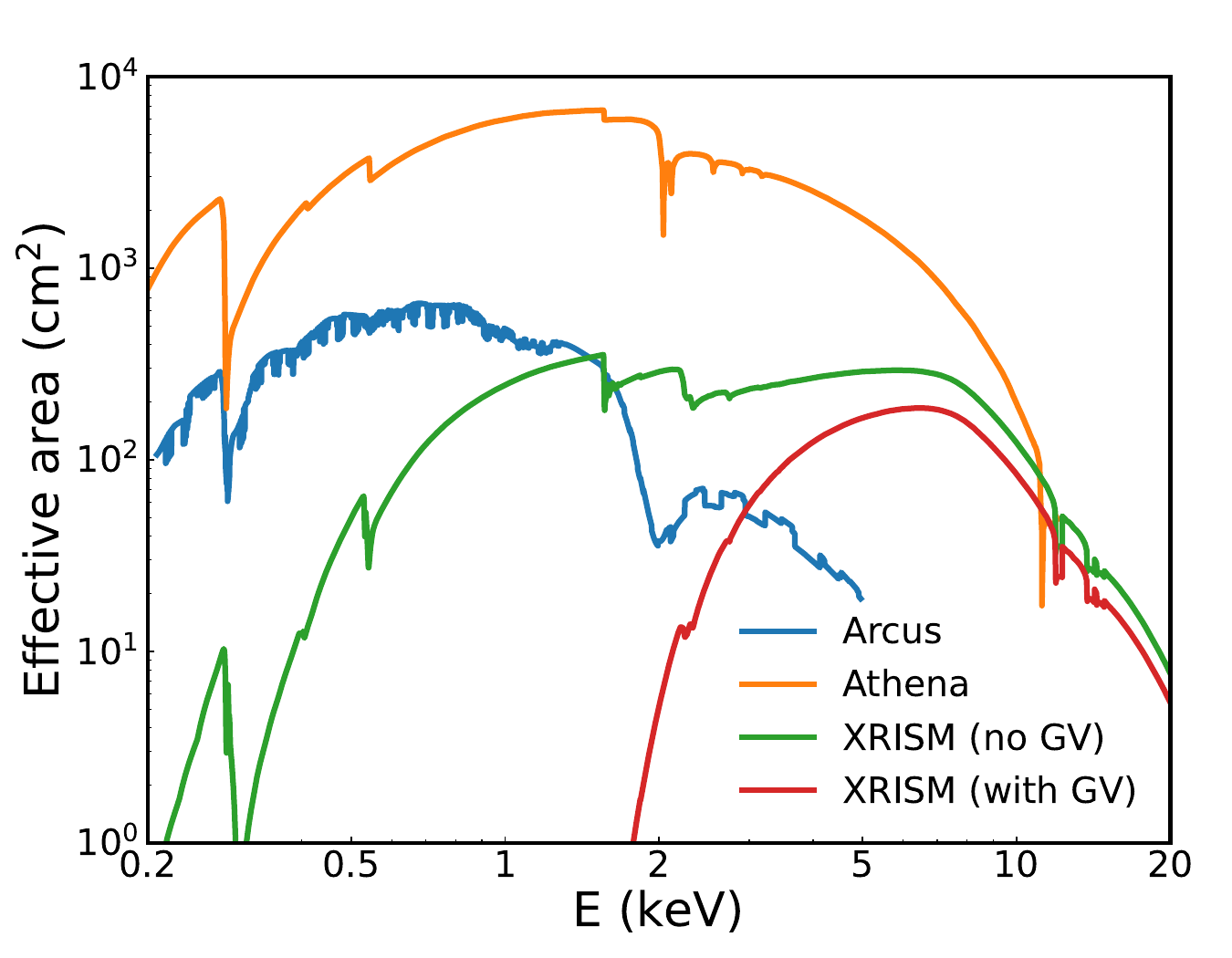}
    \caption{Effective area curves of X-ray telescopes for current and future observatories considered in this work. XRISM Resolve with the gate valve open (green) and closed (red), Athena X-IFU (orange) and Arcus XGS (blue) are shown. These curves highlight the instrumental throughput across relevant energy bands for photon–ALP conversion analysis.}
    \label{fig:effective area}
\end{figure}

In this work, we simulate the Arcus detector response using the response files\footnote{\url{http://www.arcusxray.org/simulations.html}}. The effective area curve used in our simulations is shown in Fig.~\ref{fig:effective area}.

\begin{figure*}
    \centering
\includegraphics[width=0.99\textwidth]{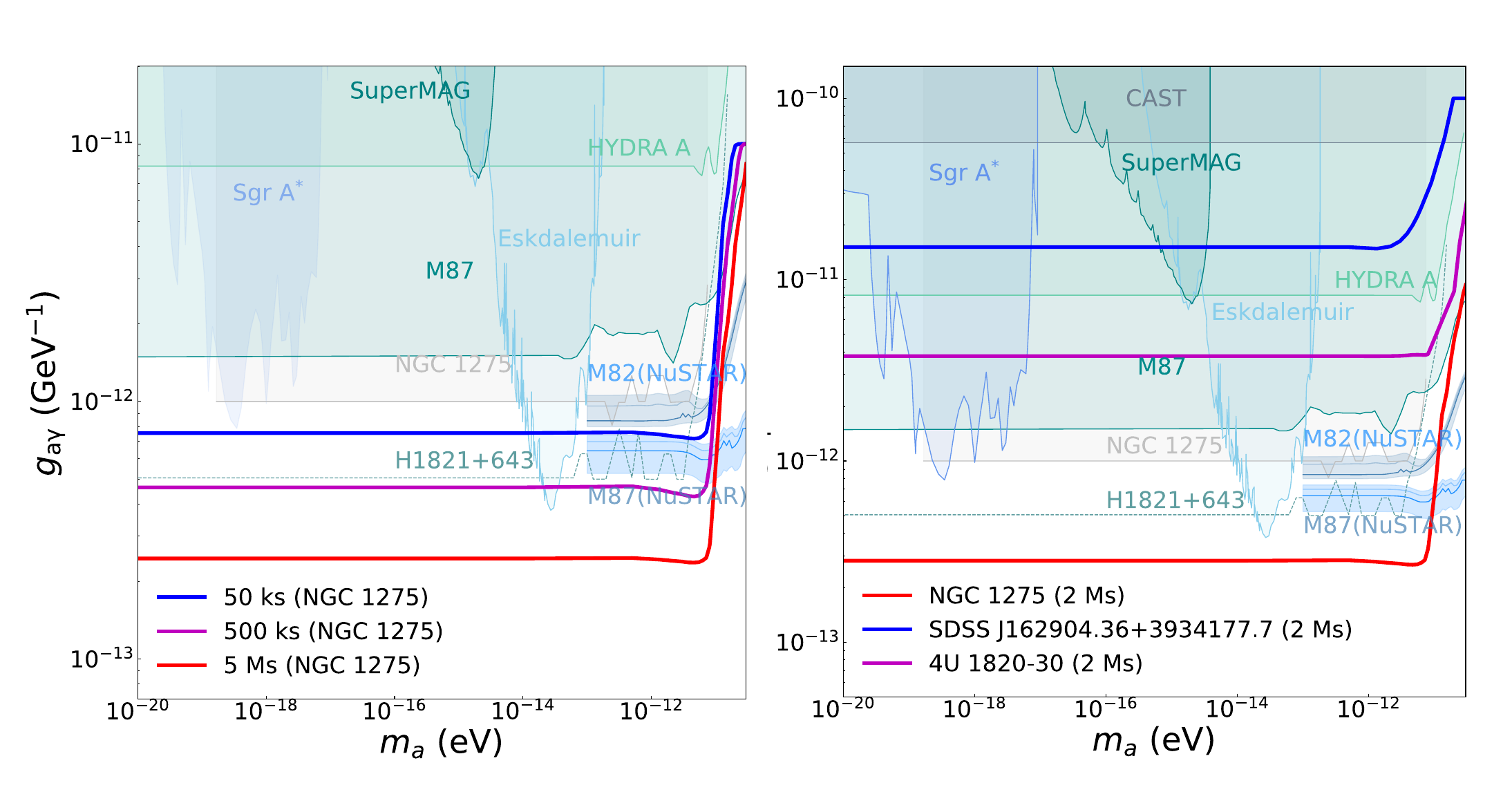}
    \caption{Sensitivity forecast for the ALP–photon coupling constant $g_{a\gamma}$ from simulated spectral fits under various observational scenarios and astrophysical target. [Left:] XRISM Resolve observations of NGC 1275 with exposure times of 50 ks, 500 ks  and 5 Ms. [Right:] 2 Ms XRISM Resolve observations of three different astrophysical systems—NGC 1275 (Perseus cluster), background quasar SDSS J162904.36+393417.7 (Abell 2199) and Galactic X-ray binary 4U 1820–30 (Milky Way). Blue shaded regions denote existing exclusion limits from previous X-ray studies reported in the literature [References: (1) Eskdalemuir~\citep{Nishizawa:2025xka}; (2) CAST-CAPP~\citep{Adair:2022rtw}; (3) SuperMAG~\citep{Arza:2021ekq,Friel:2024shg}; (4) Sgr A*~\citep{Yuan:2020xui}; (5) M87~\citep{Marsh2017}; (6) NGC 1275~\citep{Reynolds:2019uqt}; (7) H1821+643~\citep{Sisk2022}; (8) Hydra A~\citep{Wouters:2013hua}]. }  
    \label{fig:sensitivity}
\end{figure*}

\subsection{Future X-ray Detector Concepts}
\label{sec:hypo_detector}

To systematically investigate the impact of energy resolution on the sensitivity to ALP–photon coupling, we perform simulations using hypothetical detector response functions with energy resolutions of 1 eV, 10 eV  and 100 eV. These configurations span a broad range of possible future detector capabilities, from sub-eV precision to moderate-resolution instruments, allowing a controlled assessment of spectral resolution effects.

To isolate the influence of energy resolution, we assume an identical effective area across all scenarios, set equal to that of XRISM Resolve for a point source observed with the gate valve open. This is consistent with the effective area configuration of the X-ray telescope assuming no penalties in observational sensitivity from closed gate valve. The same target source spectra and observational parameters are used throughout, ensuring a consistent comparison of how energy-resolving power affects sensitivity to ALP-induced spectral modulations.

\section{Simulated Energy Spectra and Fitting Methodology}
\label{sec:fitting}

\subsection{Photon Survival Probability Model}

For each background source–galaxy cluster pair, the photon beam propagation path defining the photon–ALP conversion volume is specified as described in Sec.~\ref{sec:astrosystems} and illustrated in Fig.~\ref{fig:target}. Along these paths, we compute the photon survival probability curves using the ALPRO code, numerically solving the Schrödinger-like beam propagation equation Eq.~\eqref{equ:beam_propagation}.

The magnetic field profiles along each line of sight are modeled based on the astrophysical environment of the system considering turbulent components. The turbulent field is generated using the cell-based algorithm as reference, with cell sizes sampled from a probability distribution matched to the coherence length scales inferred for each celestial system. In App.~\ref{section:compare-B-field-models} we compare our results to the scenario of GRF magnetic field.

For each configuration, the photon survival probability is computed over 400 random turbulent field realizations. The results are averaged to obtain a representative survival curve.
As shown in Fig. \ref{fig:photon_survival_probability_mass}, the oscillation amplitudes 
in the photon survival probability curves computed from one single magnetic field are larger 
than that averaged for multiple random fields. We adopted the field-averaged photon survival 
probability models in further analyses, so that we can account for turbulent field variation 
and obtain a conservative model in terms of the final sensitivity forecast. 
These simulations are performed over a parameter space spanning ALP masses in the range $10^{-20}~\mathrm{eV} \leq m_a \leq 3\times10^{-11}~\mathrm{eV}$ and ALP–photon coupling strengths $10^{-14}~ \mathrm{GeV}^{-1} \leq g_{a\gamma} \leq 10^{-10}~\mathrm{GeV}^{-1}$.

\subsection{Target Source Spectrum Simulation}

For each background target we simulate the observed energy spectrum using the \texttt{fakeit} procedure in \texttt{XSPEC}, applying the detector response files corresponding to XRISM Resolve, Athena, Arcus and hypothetical detectors with varying energy resolution, as described in detail previously. The spectral model adopted for all the background sources is \texttt{TBabs*power-law}, where the \texttt{TBabs} component accounts for both Galactic and source's system intrinsic absorption by the interstellar medium, and the \texttt{power-law} component represents the source’s intrinsic X-ray emission. The model parameters for each system are set according to observational values described in Sec.~\ref{sec:astrosystems}. We set four control groups to investigate the impact on ALP-photon sensitivity reach by comparing: (1) the depth of exposure using XRISM Resolve 50 ks, 500 ks and 5 Ms simulations, (2) three celestial systems as the target observed by XRISM Resolve for 2 Ms and (3) different detectors, involving XRISM Resolve, Athena and Arcus considering 5 Ms observation.

\subsection{Maximum-likelihood Estimation}
\label{sec: Global Maximum-likelihood Estimation}

For all detector configurations and exposure scenarios described above we simulate 100 Monte Carlo realizations of energy spectra for each background source. These simulations incorporate the baseline astrophysical emission model modified by photon–ALP conversion effects. Specifically, the composite spectral models used are \texttt{TBabs*ALPabs*power-law} for the AGN NGC 1275 (in the Perseus cluster) and the background quasar behind Abell 2199, and \texttt{TBabs*ALPabs*(bbody+compTT)} for the Galactic X-ray binary 4U 1820–30 in the Milky Way. The \texttt{ALPabs} component represents the photon survival probability due to ALP–photon mixing and is constructed from survival curves averaged over 400 turbulent magnetic field realizations for each astrophysical environment.

Each of the 100 simulated spectra per case is fitted using the corresponding combined model. 
The astrophysical model parameters (e.g. photon index, absorption column density, and Comptonization temperatures) 
are held fixed at their input values during all fits to isolate the effect of ALP-induced 
modulations. We compute the C-statistic \citep{Cash1979}, corresponding to $ -2\ln \mathcal{L}$  
for Poisson data likelihood $\mathcal{L}$
\begin{eqnarray}
\centering
C = 2\sum_{i=1}^{N} [ t m_i - {S_i} \ln(t m_i)] ,
\end{eqnarray}
where constant terms independent of the model have been omitted, $S_i$ are the observed counts, $t$ is the observation time, 
and $m_i$ the predicted count rates based on the considered model and detector response. 


For the maximum-likelihood estimation we then minimize $C$. The C-statistic values $C(m_a, g_{a\gamma})$ 
are obtained for model parameters in the range $\rm log_{10} (m_a / eV) \in [-20, -10.5]$ and 
$\rm log_{10} (g_{a\gamma} / GeV^{-1}) \in [-13, -10] $. For each simulated realization, 
we determine the maximum value of $\Delta C(m_a, g_{a\gamma}) = C_0 - C(m_a, g_{a\gamma})$  
over the scanned parameter space $(m_a, g_{a\gamma})$. Here, $C_0$ corresponds to case without ALPs. 
We then obtain the empirical distribution $P(\Delta C_{\max})$ for $\Delta C$ from 100 
simulations. We determine the $\Delta C_{0.975}$ value where $P(\Delta C_{\max} > \Delta C_{0.975}) = 2.5\%$, 
corresponding to the one-sided 95\% confidence level. At each $m_a$, we search for the value of 
parameter $g_{a\gamma}(\Delta C_{0.975})$ where $\Delta C(m_a, g_{a\gamma}) = \Delta C_{0.975}$, 
and determine the average value of $g_{a\gamma}(\Delta C_{0.975})$ for all simulations 
as the predicted sensitivity at 95\% confidence.

This accounts for the look elsewhere effect~(see, e.g., \citet{Zhou2024}). 
Since $\Delta C_{0.975}$ is calibrated against the maximum of $\Delta C$ over
the full scanned $(m_a, g_{a\gamma})$ range rather than at a single $m_a$, this
necessarily exceeds the corresponding local threshold. The resulting projected limits
are therefore global projected upper limits with coverage $\geq 95\%$, and are
conservative relative to a local $95\%$ analysis at any individual ALP mass.

\section{Forecast for ALP-Photon Couplings}
\label{sec:forecast}

\subsection{Projected Sensitivities from NGC 1275 with XRISM/Resolve Observations}

We first examine the effect of exposure time by comparing sensitivity curves for XRISM Resolve observations of NGC 1275 with exposure times of 50 ks, 500 ks  and 5 Ms, assuming the instrument’s current gate valve–closed configuration in orbit. 95\% C.L.  projected global upper limits are shown in Fig.~\ref{fig:sensitivity}. The 50 ks XRISM exposure can achieve a sensitivity better than the 490 ks Chandra grating observation search reported in Ref.~\citep{Reynolds:2019uqt}. This highlights effects of XRISM's superior spectral resolution even with shorter integration time. Increasing the exposure to 500 ks can improve the projected limit on $g_{a\gamma}$ by approximately a factor of two. With a long 5 Ms exposure, XRISM Resolve is projected to probe couplings down to $g_{a\gamma} \simeq 2.5 \times 10^{-13}\ \mathrm{GeV^{-1}}$, demonstrating its potential to reach unexplored parameter space for ultralight ALPs.

The dependence of detection sensitivity on ALP mass can be categorized according to three characteristic regimes: massless, resonant, and massive. These correspond to cases where the ALP mass is much smaller than, comparable to, or much larger than the plasma frequency of the environment, respectively. In these regimes, the resulting photon–ALP conversion features are approximately energy-independent, resonantly enhanced  or shifted beyond the sensitive energy band, leading to distinct sensitivity behaviors. The plasma frequency range differs among the three systems we study. Approximately, the plasma frequency is $2\times 10^{-11} \sim 6\times 10^{-11}\rm\ eV$ for Milky Way and $3\times 10^{-12} \sim 1\times 10^{-11}\rm\ eV$ for Abell 2199, so the ALP mass at which resonant enhancement occurs varies accordingly across these environments. 

Below ALP mass $m_a < 5\times 10^{-13}\rm\ eV$, the detection sensitivity reaches a plateau which shows no dependence in terms of the ALP mass. This is because the ALP mass below this threshold becomes substantially small compared to the plasma frequency of the intracluster medium, which ranges between $3\times 10^{-12}\rm\ eV$ and $5\times 10^{-11}\rm\ eV$, so that the photon survival probability becomes mass-invariant compared to the photon energy in 0.5-15 keV band, which is typical sensitive range for X-ray observatories. In the mass range between $1\times 10^{-12}\rm\ eV$ and $7\times 10^{-11}\rm\ eV$, the ALP mass and plasma frequency become comparable to each other and the interacting cross section increases in a way that resonates. As a result, chances of the photon-ALP conversion's imprints on oscillating absorption features in the energy spectrum becomes deeper and the detection sensitivities are enhanced for the ALP masses in the resonance range. Beyond ALP the mass $m_a > 8\times 10^{-12}\rm\ eV$, the detection sensitivity worsens as $m_a$ increases. This is because the major oscillating absorption features moves towards higher energy range as the ALP mass increases, which gradually shifts out of the sensitive range of the detector, and the detection sensitivity of photon-ALP conversion features is lost consequently.  

We verify the statistical procedure with a signal injection test as follows. We
inject a fake ALP signal into the NGC 1275 energy spectrum with $g_{a\gamma}$
set equal to the sensitivity forecast at $m_a = 10^{-19}$~eV obtained from the
simulations without a signal. We generate $1000$ such realizations and repeat the $C$-statistic
computation of Sec.~\ref{sec: Global Maximum-likelihood Estimation} to obtain the global threshold
$\Delta C_{0.975(\mathrm{inj})}$ and the corresponding $g_{a\gamma}$
sensitivity at $m_a = 10^{-19}$~eV.

From this test, we obtain two results. First, we confirm that none of the $1000$ signal injection realizations
yield a $g_{a\gamma}$ sensitivity exceeding that of the test without signals, confirming
that our approach to estimating sensitivity is conservative. Second, comparing
the global and local thresholds directly the global critical value
$\Delta C_{0.975(\mathrm{inj})}$ lies above the $2.5\%$ critical value of
the corresponding local $\Delta C_{\mathrm{inj}}$ distribution, falling
instead at its $\sim 1.5\%$ interval. The global criterion is therefore definitively
more demanding than the local one. This arises from exclusion requiring inconsistency with
the data across the entire scanned $(m_a, g_{a\gamma})$ parameter space rather
than just at a single ALP mass. Consequently, the projected limits shown in Fig.~\ref{fig:sensitivity}
should be interpreted as global upper projected limits with coverage $\geq 95\%$, and are
conservative relative to a local $95\%$ analysis at any individual $m_a$.
Finally, adopting the criterion that a local $\Delta C_{\mathrm{inj}}$
exceeding the global $\Delta C_{0.975}$ constitutes a signal
discovery we recover a positive detection in $\sim 82\%$ of realizations,
indicating consistency of the statistical method.

\subsection{Sensitivity Scaling Relation with Exposure}

\begin{figure}
    \includegraphics[width=0.5\textwidth]{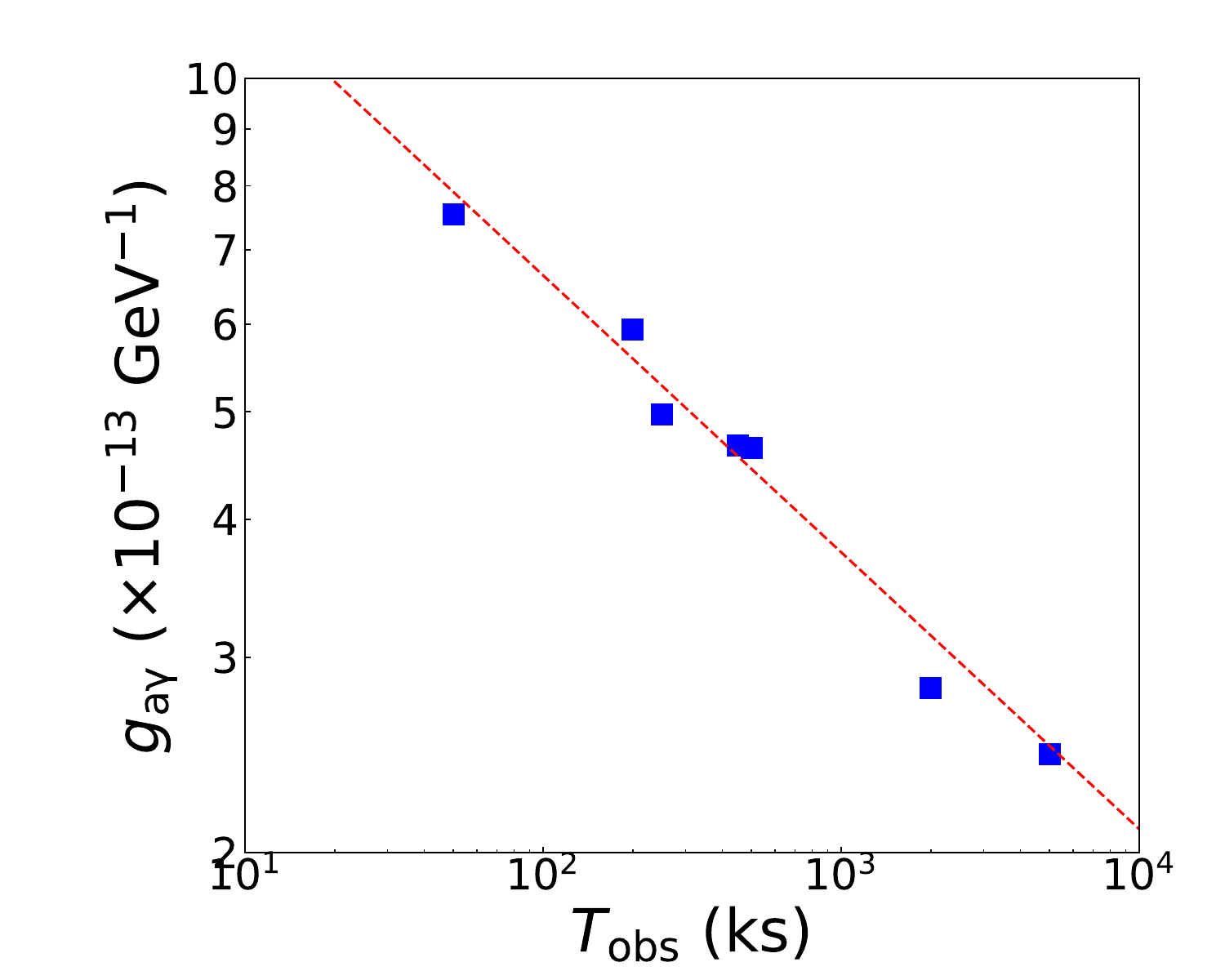}
    \caption{Dependence of the projected sensitivity (at ALP mass $m_a = 10^{-20}\rm \ eV$) to the ALP–photon coupling $g_{a\gamma}$ on XRISM/Resolve exposure time assumption, shown in blue dots. A scaling relation proportional to $T^{-1/4}$ is plotted in the red dashed line for a direct comparison.}
    \label{fig: scaling relation}
\end{figure}

To investigate the scaling relation between the projected sensitivity and the exposure time, we perform the simulation of NGC 1275 assuming a series of XRISM observation time for 50 ks, 200 ks, 250 ks, 450 ks, 500 ks, 2 Ms, and 5 Ms. The ALP detection sensitivities were searched for those simulations following the global maximum-likelihood analysis as described in Sec.~\ref{sec: Global Maximum-likelihood Estimation}. The projected sensitivities predicted at ALP mass $m_a = 10^{-20}\rm\ eV$ are plotted against the exposure time, which we find to be well approximated by $\sim T^{-1/4}$ scaling, as shown in Fig. \ref{fig: scaling relation}. From the scaling relation, we estimate that around 20 ks XRISM observation time enables ALP sensitivity to reach new parameter space beyond the previous NGC 1275 limits obtained from the Chandra grating spectrometer. It requires more than 320 ks to exceed the limit set by H1821+643, which is, however, subjected to distinct systematic uncertainties associated with  magnetic fields of the host cluster. Athena can achieve a projected sensitivity exceeding that of H1821+643 with approximately only 8 ks exposure observation of NGC 1275, assuming same sensitivity scaling relation with $T^{-1/4}$. Notably, designing an X-ray telescope with a larger effective area is equivalent to shortening the exposure time to reach the same level of photon statistics. Therefore, such scaling relationship could provide a useful guideline for the telescope design under considered scenarios.

\subsection{Impact of Magnetic Field Modeling Uncertainties on Constraints}

As shown in Fig. \ref{fig:sensitivity}, the limits set from analysis of quasar H1821+643 exceed  the previous limits from NGC 1275 by a factor of two. However, H1821+643 limits suffer  from significant uncertainties due to magnetic field modeling. In Ref.~\citep{Sisk2022}, a fiducial magnetic field model with a radial profile similar to Perseus cluster was adopted, assuming near 9 $\mu$G  B-field at $\sim 10\rm\ kpc$ and $1\sim 5\rm\ \mu G$ within $10\sim 100\rm\ kpc$. This is a significant assumption for ALP-mixing modeling, given that the host cluster CL1821+643 has no Faraday rotation measures available for solid field modeling. According to the radio synchrotron measurement of the source and standard synchrotron description, the estimated field is about 0.3 $\mu \rm G$ if the radio halo is uniform and has unity filling factor throughout an ellipsoidal volume. Further, there are additional assumptions  such as equal energy density between the relativistic protons and electrons, and minimization of the particle and magnetic field energy density at the observed synchrotron emissivity. To bring the inferred field strength to higher value, i.e. 2 $\mu \rm G$ at 500 kpc, enhancing of the relativistic proton-to-electron energy density ratio to $\sim 50$ or even higher is required. Yet another significant uncertainty arises from the geometry of the radio relic CL1821+643, and viewing line of sight edge-on or face-on can significantly affect the filling factor and the inferred magnetic field strength. Therefore, we consider that the limit from H1821+643 has distinct and notable uncertainties compared to  NGC 1275, necessitating other robust observations and searches in such parameter space.

\subsection{Comparisons among Different Target Systems}

A comparison of different astrophysical targets under identical observing conditions considering 2 Ms exposure with XRISM Resolve reveals significant variation in sensitivity driven by both source flux and magnetic field structure. Among the three systems analyzed, NGC 1275 in the Perseus cluster emerges as the most favorable target, offering the strongest projected sensitivity reach to the photon–ALP coupling $g_{a\gamma}$. The background quasar SDSS J162904.36+393417.7  located behind the Abell 2199 cluster yields a sensitivity projections that are approximately two orders of magnitude weaker, primarily due to its significantly lower X-ray flux. In contrast, the Galactic X-ray binary 4U 1820–30 is intrinsically bright and thus not limited by photon statistics. However, its relatively short path length through the Galactic magnetic field and the smaller conversion volume resulting from its proximity lead to a higher sensitivity floor compared to the extragalactic cluster target.

Simulation of the background quasar Abell 2199 suggests that the ALP detection sensitivity reach of such system, $\sim 1.5\times 10^{-11}\rm\ GeV^{-1}$ at ALP mass of $m_a = 10^{-20}\rm \ eV$, is significantly lower than from a search with NGC 1275. We consider if   stacking many of such background samples can improve the ALP detection sensitivity. For this, we estimate possible enhancement of sensitivity assuming collection of all the background AGN sources whose fluxes are larger than $10^{-13}\rm\ erg\ cm^{-2}\ s^{-1}$ throughout the whole sky. The number density of such sources is found to be around ten per degree square according to the $\ln N - \ln S$ relation of the X-ray sources in 2-10 keV band, which yields a total number of about $4\times 10^5$ for $4\pi$ solid angle. We find that stacking of all such sources improves the total statistics by a factor of $4\times 10^5$, equivalent to increase the exposure time by a similar factor. Since not all background sources pass through a magnetic field volume   along each line of sight  this estimate only provides an upper limit. Following the discussed above sensitivity-exposure scaling relation, we expect that the ALP sensitivity reach with such approach can be enhanced from $1.5\times 10^{-11} \ \rm GeV^{-1}$ to $\sim 6.7\times 10^{-13} \ \rm GeV^{-1}$, which can surpass the sensitivity from analysis of NGC 1275 by approximately a factor of two. Hence, stacking the background quasar signals could yield results competitive with ALP-photon mixing search analyses of NGC 1275 and Perseus cluster system.

One can consider searching for X-ray sources brighter than NGC 1275. Given that NGC 1275 is already the brightest among all AGN residing in the center of the galaxy cluster, Galactic X-ray source systems of the Milky Way constitute among the best alternatives that could provide a background target that is both brighter and closer to the Earth. The brightest galactic X-ray source, Sco X-1, is approximately thirty times brighter than NGC 1275. However, Sco X-1 brightness can  saturate the detector in case of direct observations. Therefore, we consider 4U 1820-30 system that is among the brightest Galactic X-ray sources and relatively distant to the Earth, which enables to consider a sizable effective magnetic field volume relevant for ALP-photon mixing analysis. We analyze ALP detection sensitivity obtained from 4U 1820-30 simulations, which is found to be $\sim 4\times 10^{-12} \ \rm GeV^{-1}$ at an ALP mass of $m_a = 10^{-20}\rm \ eV$. Thus, one would need to analyze a combined stack of signals from $\sim 10^4$ Milky Way X-ray sources within similar fluxes in order to reach a sensitivity comparable to NGC 1275. However, there are only several dozens of persistent bright Galactic X-ray binary system sources known thus far. This highlights that NGC 1275 represents a compelling target for ALP analysis compared to other astrophysical systems and understanding its systematics and those of Perseus cluster itself play an important role.

\begin{figure}
    \centering
    \includegraphics[width=0.5\textwidth]{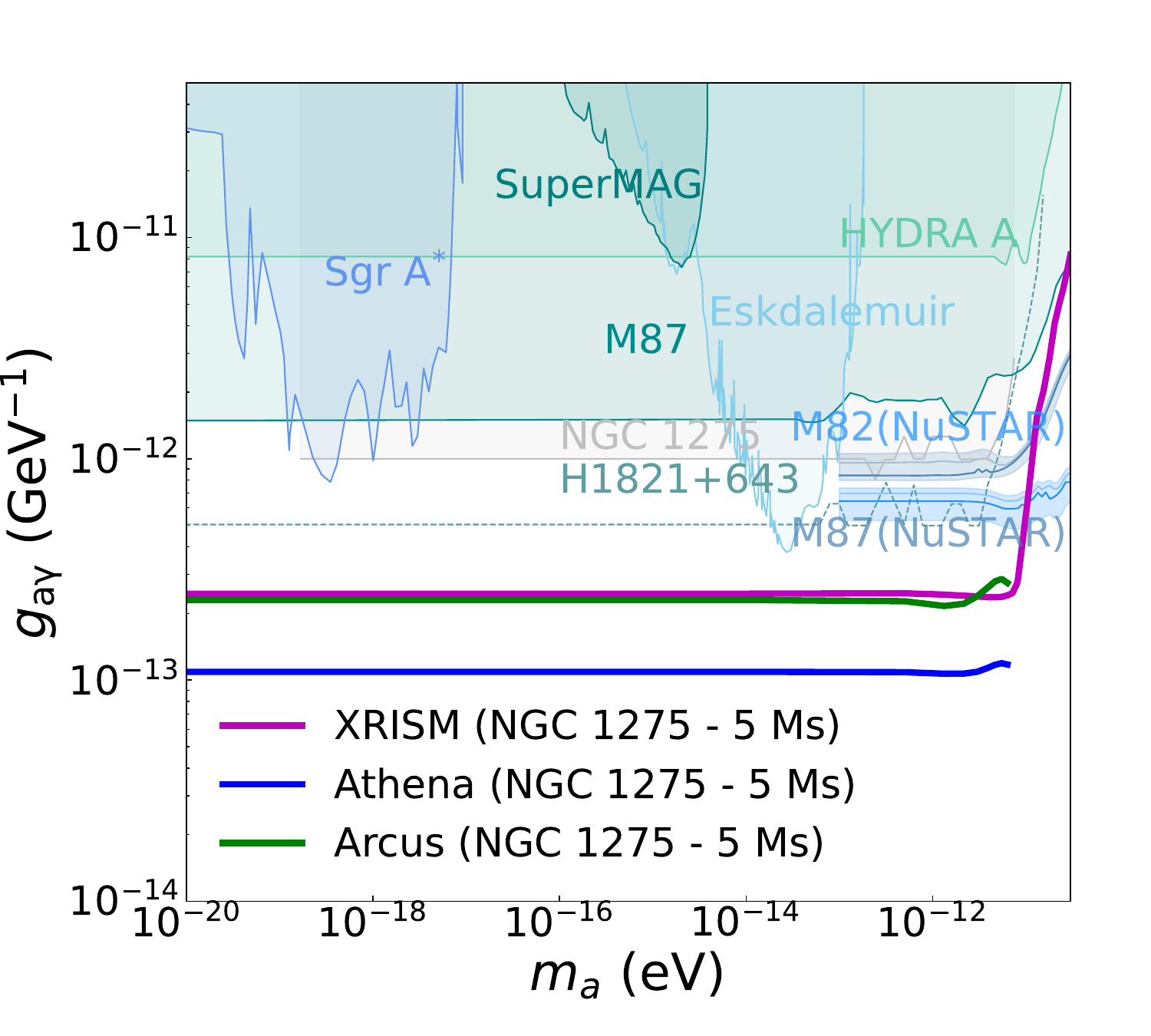}
    \caption{5 Ms observations of NGC 1275 using the instrument responses of XRISM Resolve, Athena X-IFU  and Arcus XGS. Sensitivity curves represent the projected mean 95\% global upper limits searched in the ALP mass range $10^{-20}\rm\ eV \leq m_a \leq 3\times 10^{-11}\rm\ eV$. }
    \label{fig:observatory}
\end{figure}

\subsection{Forecasts for Future missions}

We also assess the potential of next-generation X-ray observatories, including Athena and Arcus, in comparison with XRISM Resolve. The projected sensitivities are shown in Fig.~\ref{fig:observatory}. Fig.~\ref{fig:effective area} displays the corresponding effective area curves, illustrating the instrumental throughput across the relevant energy range. For a fixed 2 Ms exposure, Athena’s X-IFU is expected to achieve up to a factor of two improvement in sensitivity over XRISM Resolve for ALP masses below $m_a \lesssim 10^{-12}$ eV. This enhanced performance arises primarily from Athena’s larger effective area and superior energy resolution in the soft X-ray band, which together improve both photon statistics and sensitivity to ALP-induced spectral modulations.

The effective area of both Arcus and Athena dominates only below 5 keV, which limits their performance in detecting ALP–photon conversion features at higher ALP masses. Consequently, we present sensitivity curves for Arcus and Athena only up to $6\times 10^{-12}$~eV.
For Arcus the sensitivity is not limited by energy resolution, but rather by the highly structured and non-uniform effective area as a function of energy that complicates the identification of the quasi-oscillatory spectral modulations expected from ALP–photon mixing. To mitigate this, we apply a coarser spectral binning scheme requiring at least $10^4$ counts per bin, which effectively smooths over artificial instrumental features. With this approach, we find that Arcus can achieve sensitivity comparable to XRISM Resolve in the 2–5 keV range, where the effective areas of both instruments are similar. This highlights that while Arcus offers excellent spectral resolution, its utility for ALP searches is influenced by its throughput profile and energy-dependent response.

\begin{figure}
    \centering
    \includegraphics[width=0.5\textwidth]{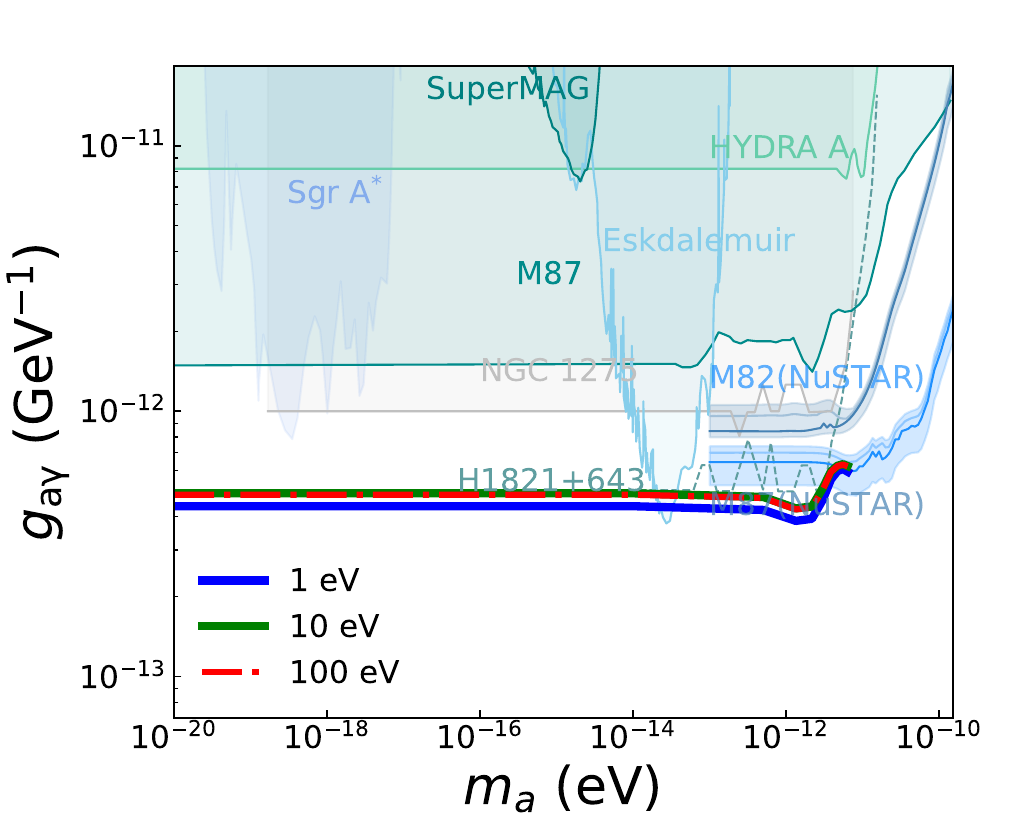}
    \caption{Projected ALP-photon mixing coefficient sensitivity curves forecast for the hypothetical detectors assuming energy resolutions of 1 eV, 10 eV, and 100 eV. 250 ks exposure time observations were simulated with XRISM/Resolve effective area at the gate valve open status.}
    \label{fig:resolution}
\end{figure}

\subsection{Effects of Energy Resolution and Binning}

In order to optimize future ALP searches, we investigate how the detector's energy resolution affects the sensitivity forecasts, in conjunction with analysis binning employed in spectral fitting. While high energy-resolving power enhances the ability to resolve fine spectral features, such as the oscillatory modulations induced by photon-ALP mixing, it also results in lower photon statistics per energy bin. This trade off may degrade or enhance the signal-to-noise (SNR) ratio.

To quantify these effects we simulate hypothetical detector responses with energy resolutions of $\Delta E = 1~\mathrm{eV}$, $10~\mathrm{eV}$ and $100~\mathrm{eV}$ that could be representative of superconducting microcalorimeters, semiconductor detectors and CCD-type detectors envisioned in next-generation X-ray observatories.  Considering response functions for hypothetical detectors (see Sec.~\ref{sec:hypo_detector}), we simulate observations for a 250 ks exposure using the \texttt{XSPEC} routine \texttt{fakeit}, producing 1000 realizations for each configuration.

We subsequently modify the binning of the simulated spectra such that each energy bin contains at least a specified number of total photon counts ranging from 100 to 3000. For each binned spectrum  we apply a combined model incorporating both the astrophysical background continuum spectrum and ALP-photon conversion effects. We  perform spectral fits to derive projected confidence limits on the ALP-photon coupling constant $g_{a\gamma}$ for a given range of ALP mass $m_a$. 

We find that detectors with 1 eV resolution can achieve the best ALP coupling $g_{a\gamma}$ sensitivity, reaching $\sim 4.4\times 10^{-13}\ \rm GeV^{-1}$ at ALP masses of $m_a = 10^{-20}\rm \ eV$, while detectors with 10 eV and 100 eV resolution show similar sensitivity levels of $\sim 4.9\times 10^{-13}\ \rm GeV^{-1}$. 
For a fixed ALP mass $m_a$, we find that the local minimization of the C-statistic is insensitive to the choice of spectral binning. In particular, rebinning the energy channels does not degrade the sensitivity, indicating that the relevant information is effectively preserved and that the sensitivity saturates once the spectral features are sufficiently resolved. Consequently, detectors with finer intrinsic energy resolution do not provide additional gains beyond this point. In contrast, when deriving global 95\% C.L. sensitivity over the full ALP mass range (as described in Sec.~\ref{sec: Global Maximum-likelihood Estimation}), we observe a slight degradation in the final sensitivity when adopting a fixed binning of 1 eV. This reflects a modest loss of information associated with discretization across the full parameter scan.
We find that the differences in final sensitivities are no more than $\sim 1\%$ for 1 eV resolution detectors among the data binned with group counts ranging from 100 to 3000. All data grouped for this range by detectors with 1-100 eV resolutions show a $g_{a\gamma}$ sensitivity consistent within $\mathcal{O}(10) \%  $.

\section{Conclusions}
\label{sec:conclusions}

High-resolution X-ray spectroscopy with upcoming and future telescopes offers a powerful avenue to probe physics beyond the SM, especially in searches for ALPs. Using detailed simulations, we have comprehensively analyzed prospects for detecting ALP-photon conversion signatures across several representative astrophysical systems consisting of AGNs in galaxy clusters, background quasars viewed through foreground clusters and Galactic X-ray binaries.

We have systematically examined how instrumental parameters, magnetic field modeling and analysis choices 
impact resulting sensitivity for ALP detection. We found that the projected sensitivity scales to a good 
approximation as $T^{-1/4}$ with exposure time $T$. We identify NGC 1275 as a particularly compelling 
system for ALP searches, compared to background quasars viewed through foreground clusters and Galactic X-ray binaries. 
A robust understanding of the associated uncertainties in modeling of NGC 1275 and the Perseus cluster 
plays a prominent role in the ALP-photon mixing study. 

Our analysis demonstrates that observations of AGNs with XRISM, Athena and Arcus can access previously unexplored regions of ALP parameter space for masses $m_a \lesssim 10^{-11}\mathrm{eV}$. 
Based on our simulations and the scaling of projected ALP-photon coupling sensitivity, we find that around $\sim$20 ks XRISM/Resolve observation of NGC 1275 can allow surpassing previous limits 
set by Chandra grating observations, 
while an exposure of $\sim320$~ks is required to exceed the sensitivity compared to constraints from H1821+643.
For Athena, the enhanced effective area enables a comparable sensitivity to be reached with only $\sim8$ ks exposure, demonstrating the potential advantage of next generation instruments.

Our results demonstrate that high-resolution X-ray observations when combined with optimized target selection and robust statistical analysis can significantly enhance sensitivity to ultralight ALPs. Future X-ray mission such as Athena offers promising opportunities to probe previously inaccessible regions of ALP parameter space. These searches complement laboratory and other astrophysical probes together spanning a broad region of ALP parameter space and thus offer a powerful probe of new physics beyond SM.

\acknowledgments

This work was supported by the World Premier International Research Center Initiative (WPI), MEXT,
Japan. V.T. acknowledges support from JSPS KAKENHI grant. No. 23K13109, and K.M. No. 20H05857.

\appendix

\section{Impact of Magnetic Field Modeling on ALP-Photon Sensitivity Forecasts}
\label{section:compare-B-field-models}

\begin{figure}[t]
    \centering
    \includegraphics[width=0.5\textwidth]{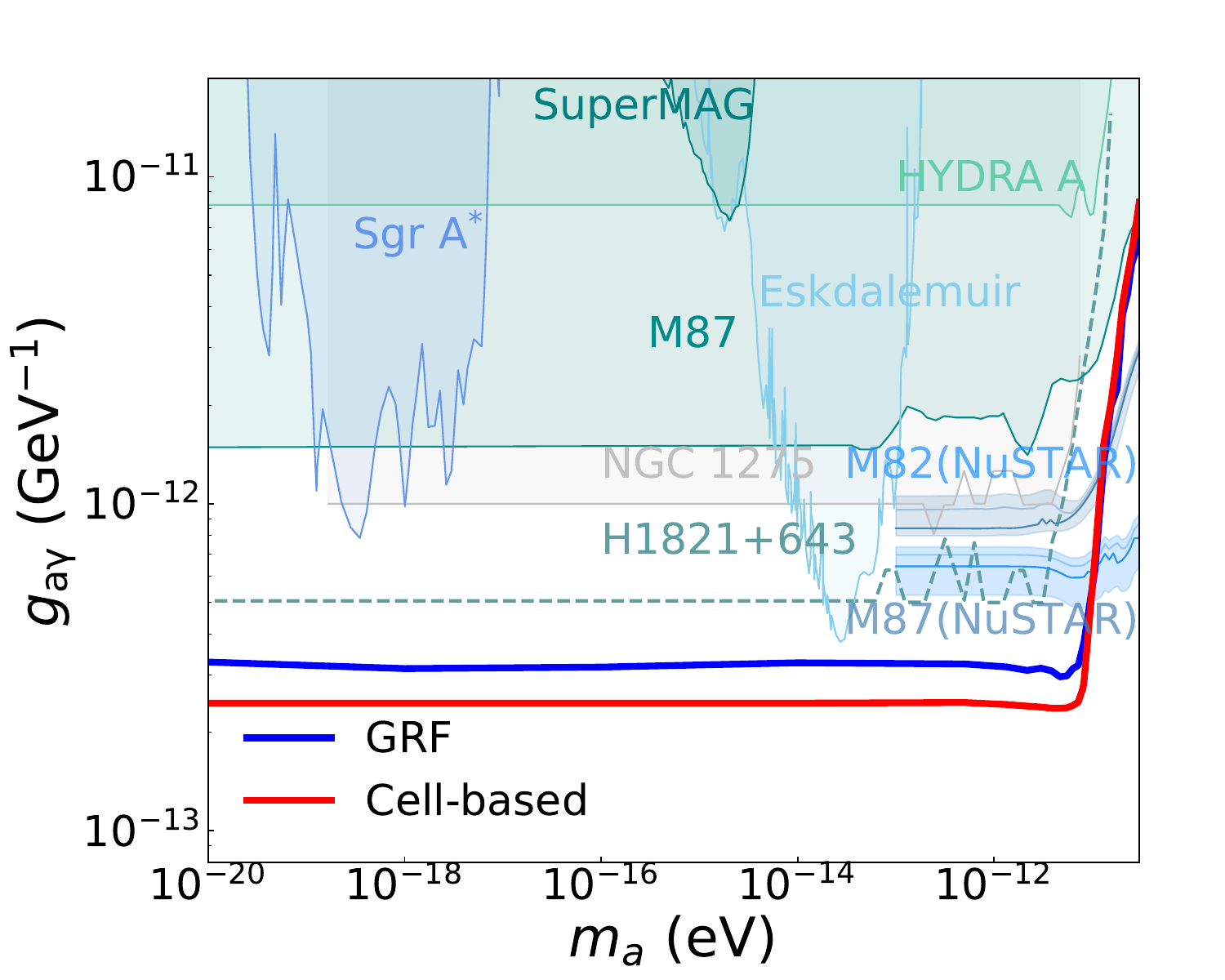}
    \caption{Impact of the turbulent magnetic field model on the sensitivity to the ALP–photon coupling constant $g_{a\gamma}$, based on simulations of a 5 Ms XRISM Resolve observation targeting NGC 1275. The red curve corresponds to simulations using the cell-based magnetic field algorithm and the blue curve uses the GRF algorithm. Both models are normalized to match the same radial magnetic field profile. The comparison illustrates the influence of different turbulence modeling approaches on the resulting sensitivity forecasts.}
    \label{fig:sensitivity-Bfield}
\end{figure}

To investigate how the choice of modeling algorithm for simulating turbulent magnetic fields affects the projected sensitivity to ALP–photon coupling we compare the cell-based algorithm described in Sec.~\ref{sec:cell-based} and GRF method introduced in Sec.~\ref{sec:GRF}. We consider the NGC 1275 Perseus cluster system as a representative case. We assume a 5 Ms XRISM Resolve observation for this comparison.
Both methods adopt the same underlying radial magnetic field profile, specified in Eq.~\eqref{equ : B-field}, as the normalization baseline. For the GRF case, we generate 400 realizations of the turbulent magnetic field and compute the corresponding photon survival probability curves from ALP–photon conversion. The mean survival probability is then derived and used to fit simulated energy spectra. To assess the sensitivity, we simulate 100 spectra for the background source and apply the fitting procedure to extract the 95\% C.L. global upper limit on the coupling constant $g_{a\gamma}$. Fig.~\ref{fig:sensitivity-Bfield} presents the resulting $g_{a\gamma}$ sensitivity forecasts for different magnetic field modeling. We find that the GRF-generated field yields slightly weaker sensitivity projections compared to the cell-based approach, which can be attributed to higher variability in the GRF turbulent magnetic field realizations.

\section{XMM-Newton Data Reduction for Background Quasar Catalog}
\label{app:secqso}

In order to identify optimal background quasar target sources we perform spectral and imaging analysis using the XMM-Newton Science Analysis System\footnote{\url{https://www.cosmos.esa.int/web/xmm-newton/sas}}. Event files for the two onboard cameras, EPIC-pn and EPIC-MOS, are generated using the standard procedures \texttt{epchain} and \texttt{moschain}, respectively. To remove time intervals contaminated by soft proton flares, we apply the \texttt{pn-filter} and \texttt{mos-filter} routines. Based on light curve inspections, all flaring intervals are excluded from further analysis.

Since the quasars exhibit negligible spatial extent on the EPIC detectors, we treat them as point sources for spectral extraction. Events within the source region are selected using \texttt{evselect}, with standard filtering criteria applied: ``\texttt{(XMMEA\_EM) \&\& (PATTERN$\leq$12)}'' for EPIC-MOS and ``\texttt{(FLAG==0) \&\& (PATTERN$\leq$4)}'' for EPIC-pn. The associated response matrix file and ancillary response file are generated using \texttt{rmfgen} and \texttt{arfgen}, respectively.

For background subtraction we extract spectra from a nearby blank sky region using the same procedures. The source and background extraction regions are normalized using \texttt{backscale}. We subtract the scaled background spectrum from the source spectrum and group the resulting channels to ensure a minimum of 25 counts per bin, which enables the application of $\chi^2$ statistics in spectral fitting.

Tab.~\ref{tab:qso_info} summarizes the properties of the quasar–galaxy cluster pairs considered in our analysis. 
The XMM-Newton observations used to extract the background quasar spectra are listed in Tab.~\ref{tab_catalog}.

\addtocounter{table}{1}  
\renewcommand{\thetable}{\arabic{table}}  

\begin{deluxetable*}{lcccccccc}[t]

\label{tab:qso_info}
\tablenum{2}
\tablewidth{0pt}
\tabletypesize{\normalsize}
\tablecaption{ Summary of the background QSO and foreground galaxy cluster (GlC) pairs. The columns are (1) QSO index, (2) QSO name, (3) QSO right ascension (J2000), (4) QSO declination (J2000), (5) QSO redshift, (6) observed X-ray flux of the QSO in the $0.3-10~\rm keV$ bands in logarithm, (7) name of the galaxy clusters (GlC), (8) redshift of the galaxy clusters, (9) averaged central magnetic field of the galaxy clusters. References: 1.~\citet{Abazajian2009_qso1z}, 2. \citet{Alam2015_qso2}, 3. \citet{Hewitt1989_qso4}, 4. \citet{Ai2016}, 5. \citet{SloanSurvey2018}, 6. \citet{2020ApJS..250....8L}, 7. \citet{2012ApJS..203...21A}, 8. \citet{Falco1999_3c449z}, 9. \citet{2015ApJS..219...18X}, 10. \citet{Rines2016_A1656z}, 11. \citet{Struble1987_A665zA2634za194z}, 12. \citet{Bonafede2010_coma}, 13. \citet{Murgia2004_a119}, 14. \citet{Vacca2012}, 15. \citet{Vogt2003_a2634}, 16. \citet{Govoni2017_a194}, 17. \citet{Guidetti2010_3c449}, 18. \citet{Vacca2010_a665}. }
\tablehead{
\colhead{srcID} & \colhead{QSO Name} & \colhead{RA} & \colhead{Dec} & \colhead{$z_{\rm QSO}$} & \colhead{$\log F_{\rm X, 0.3-10}$} & \colhead{GlC Name}  & $z_{\rm GlC}$ & \colhead{GlC B-field} \\
\colhead{} & \colhead{} & \colhead{(deg)} & \colhead{(deg)} & \colhead{} & \colhead{$(\rm erg~cm^{-2}~s^{-1})$} & \colhead{}  & \colhead{} & \colhead{$(\rm \mu G)$} 
}
\startdata
 \hline
 1 & SDSS J125859.27+275308.5 & 194.7470 & 27.8860 & $1.135$ $^1$ & $-13.22^{+0.13}_{-0.17}$  & \multirow{7}{*}{A1656} & \multirow{7}{*}{0.0234 $^{10}$} & \multirow{7}{*}{$3-7^{12}$} \\
 2 & SDSS J125915.61+280804.8 & 194.8150 & 28.1347 & $2.078$ $^2$ & $-13.51^{+0.33}_{-0.17}$ &   &   &   \\
 3 & SDSS J125831.74+275330.2 & 194.6323 & 27.8917 & $1.141$ $^2$ & $-13.19^{+0.11}_{-0.08}$ &   &   &   \\
 4 & QSO B1258+280 & 195.2942 & 27.8164 & 1.930 $^3$ & $-13.17^{+0.03}_{-0.03}$ &   &   &   \\
 5 & 2XMM J125926.2+282332 & 194.8592 & 28.3922 & 1.154 $^4$ & $-12.77^{+0.01}_{-0.02}$ &   &   &   \\
 6 & QSO B1258+2835 & 195.2536 & 28.3291 & $1.361$ $^2$ & $-12.93^{+0.02}_{-0.02}$ &   &   &   \\
 7 & QSO B1258+2839 & 195.2004 & 28.3891 & $1.924$ $^2$ & $-12.86^{+0.01}_{-0.02}$ &   &   &   \\
 \hline
 8 & SDSS J005606.43-011958.4  & 14.0268 & -1.3329 & $0.840$ $^5$ & $-13.35^{+0.04}_{-0.05}$ & \multirow{1}{*}{A119} & \multirow{1}{*}{0.0445 $^{10}$} & \multirow{1}{*}{$5^{13}$}  \\
 \hline
 9 & SDSS J162829.75+392115.1 & 247.1240 & 39.3542 & 0.383 $^9$ & $-12.9^{+0.03}_{-0.04}$ & \multirow{7}{*}{A2199} & \multirow{7}{*}{0.0309 $^{10}$} & \multirow{7}{*}{$5^{14}$} \\
 10 & [VV2006] J162855.6+394034 & 247.2317 & 39.6761 & $1.520$ $^2$ & $-12.88^{+0.05}_{-0.03}$ &   &   &   \\
 11 & SDSS J162806.89+393911.0 & 247.0287 & 39.6531 & 1.164 $^6$ & $-13.47^{+0.11}_{-0.11}$ &   &   &   \\
 12 & SDSS J162809.81+392814.7 & 247.0409 & 39.4708 & 1.159 $^6$ & $-13.23^{+0.03}_{-0.03}$ &   &   &   \\
 13 & SDSS J162904.36+393417.7 & 247.2682 & 39.5716 & 0.956 $^6$ & $-12.71^{+0.01}_{-0.02}$ &   &   &   \\
 14 & SDSS J162906.84+393240.5 & 247.2785 & 39.5446 & 0.783 $^5$ & $-12.81^{+0.02}_{-0.02}$ &   &   &   \\
 15 & [VV2006] J162937.1+394100 & 247.4046 & 39.6833 & $0.724$ $^7$ & $-12.93^{+0.03}_{-0.03}$&   &   &   \\
 \hline
 16 & SDSS J233831.66+270035.0 & 354.6320 & 27.0097 & 1.649 $^6$ & $-13.12^{+0.08}_{-0.09}$ & \multirow{2}{*}{A2634} & \multirow{2}{*}{0.0312 $^{11}$} &  \multirow{2}{*}{$11.7\pm9.0^{15}$} \\
 17 & SDSS J233820.00+265638.7  & 354.5834 & 26.9441 & 1.289 $^6$ & $-13.01^{+0.07}_{-0.07}$ &   &   &   \\
 \hline
 18 & SDSS J012555.11-012925.1 & 21.4796 & -1.4903 & 2.658 $^5$ & $-14.18^{+1.01}_{-0.37}$ & \multirow{4}{*}{A194} & \multirow{4}{*}{0.0178 $^{11}$} & \multirow{4}{*}{$1.5\pm0.2^{16}$} \\
 19 & SDSS J012556.54-012731.8 & 21.4856 & -1.4588 & $2.808$ $^7$ & $-14.09^{+0.54}_{-0.28}$ &   &   &   \\
 20 & SDSS J012610.47-012605.0 & 21.5436 & -1.4347 & $1.063$ $^5$ & $-13.62^{+0.09}_{-0.41}$ &   &   &   \\
 21 & SDSS J012556.71-012607.7  & 21.4863 & -1.4355 & $1.887$ $^5$ & $-14.03^{+0.43}_{-0.24}$ &   &   &   \\
 \hline
 22 & UGC12064 & 337.8358 & 39.3582 & $0.017$~$^8$ & $-12.92^{+0.01}_{-0.09}$ & \multirow{1}{*}{3C449} & \multirow{1}{*}{$0.0171$~$^8$} & \multirow{1}{*}{$3.5\pm1.2^{17}$}  \\
 \hline
 23 & 2XMM J083107.3+654653 & 127.7804 & 65.7814 & 0.638 $^9$ & $-12.80^{+0.02}_{-0.02}$ & \multirow{1}{*}{A665} & \multirow{1}{*}{0.0186 $^{11}$} & \multirow{1}{*}{$1.3^{18}$} \\
\enddata
\end{deluxetable*}

\addtocounter{table}{1} 

\begin{deluxetable*}{lccc}[t]

\label{tab_catalog}
\tablenum{3}
\tablewidth{0pt}
\tabletypesize{\normalsize}
\tablecaption{Summary of the \textit{XMM-Newton} observations used for the QSO–galaxy cluster pairs. The columns are (1) observation ID, (2) observation start date, (3) exposure time in kiloseconds (ks), (4) QSO index (see Tab.~\ref{tab:qso_info}) to which each observation applies.  }
\tablehead{
\colhead{Obs.ID} & \colhead{Start date} & \colhead{Duration time} & \colhead{srcID}  \\
\colhead{} & \colhead{} & \colhead{(ks)} & \colhead{} 
}
\startdata
 0124710501 & 2000-05-29 & 26.0 & 1, 3 \\
 0124710601 & 2000-06-12 & 18.2 & 1, 2, 3 \\
 0124710801 & 2000-12-10 & 27.6 & 4 \\
 0124710901 & 2000-06-11 & 30.2 & 4 \\
 0124711401 & 2000-05-29 & 23.7 & 1, 2 \\
 0124712001 & 2000-12-10 & 21.9 & 1, 2, 3 \\
 0124712501 & 2002-06-07 & 28.0 & 6 \\
 0153750101 & 2001-12-04 & 25.2 & 1, 2 \\
 0204040101 & 2004-06-06 & 87.9 & 5, 6, 7 \\
 0204040201 & 2004-06-18 & 101.5 & 5, 6, 7 \\
 0204040301 & 2004-07-12 & 98.1 & 5, 6, 7 \\
 0300530101 & 2005-06-18 & 25.3 & 1 \\
 0300530201 & 2005-06-17 & 15.9 & 1, 2 \\
 0300530301 & 2005-06-11 & 30.8 & 1 \\
 0300530401 & 2005-06-09 & 27.3 & 1 \\
 0300530501 & 2005-06-08 & 25.3 & 1 \\
 0300530601 & 2005-06-07 & 25.5 & 1, 2, 3 \\
 0300530701 & 2005-06-06 & 25.3 & 1, 2 \\
 0304320201 & 2005-06-28 & 79.8 & 5, 6, 7 \\
 0304320301 & 2005-06-27 & 55.8 & 5, 6, 7 \\
 0304320801 & 2006-06-06 & 63.6 & 5, 6, 7 \\
 0851180501 & 2019-05-30 & 47.0 & 4 \\
 0864560101 & 2020-07-11 & 133.5 & 4 \\
 0904640201 & 2022-06-18 & 103.2 & 4 \\
 0904640301 & 2022-06-19 & 113.4 & 4 \\
 0402190501 & 2006-06-16 & 19.7 & 8 \\
 0505211001 & 2007-07-14 & 13.6 & 8 \\
 0012440101 & 2001-01-14 & 29.2 & 8 \\
 0760340201 & 2015-07-06 & 91.6 & 8 \\
 0136340101 & 2002-12-23 & 22.3 & 18, 19, 20, 21 \\
 0743700201 & 2015-01-10 & 50.2 & 18, 19, 20, 21 \\
 0002960101 & 2002-06-22 & 11.3 & 16, 17 \\
 0505210801 & 2007-12-26 & 15.0 & 16, 17 \\
 0800761501 & 2017-07-10 & 8.5 & 16, 17 \\
 0002970101 & 2001-12-09 & 21.1 & 22 \\
 0723801101 & 2013-08-12 & 55.6 & 9, 10, 11, 12, 13, 14, 15 \\
 0723801201 & 2013-08-29 & 54.1 & 9, 10, 11, 12, 13, 14, 15 \\
 0008030301 & 2002-07-06 & 13.3 & 9, 10, 11, 12, 13, 14, 15 \\
 0008030601 & 2002-08-15 & 8.1 & 9, 10, 11, 13, 14, 15 \\
\enddata
\end{deluxetable*}

\bibliographystyle{aasjournal}
\bibliography{refs} 

\end{document}